\documentclass{aa}  
\usepackage{graphicx}
\usepackage{txfonts}
\usepackage{natbib}
\bibpunct{(}{)}{;}{a}{}{,}
\usepackage{aalongtable, lscape}
\begin{document}
   \title{The stellar content of the XMM-Newton Bright Serendipitous Survey
   \thanks{The XMM-Newton Bright
    Serendipitous Survey was conceived by the XMM-Newton Survey Science Center
    (SSC), a consortium of 10 institutions appointed by ESA to help the SOC in
    distinct technical aspects, including the exploitation of the XMM-Newton
    serendipitous detections (see http://xmmssc-www.star.le.ac.uk/).}}

   \subtitle{}

   \titlerunning{The stellar content of the XBSS}
   \authorrunning{J. L\'opez-Santiago et al.}

   \author{J. L\'opez-Santiago\inst{1}
          \and
          G. Micela\inst{1}
          \and
          S. Sciortino\inst{1}
          \and
          F. Favata\inst{2}
          \and
          A. Caccianiga\inst{3}
          \and
          R. Della Ceca\inst{3}
          \and
          P. Severgnini\inst{3}
          \and
          V. Braito\inst{4, 5}
          }

   \offprints{J. L\'opez-Santiago,\\ \email{jlopez@astropa.unipa.it}}

   \institute{INAF - Osservatorio Astronomico di Palermo 
              Giuseppe S. Vaiana, 
              Piazza Parlamento 1, I-90134 Palermo, Italy\\
              \email{jlopez@astropa.unipa.it}
         \and
              Astrophysics Missions Division, Research and Science Support 
              Department of ESA/ESTEC, Postbus 299, 2200 AG Noordwijk, 
              Netherlands
         \and
              INAF - Osservatorio Astronomico di Brera,
              via Brera 28, I-20121 Milan, Italy
         \and
              X-ray Astrophysics Laboratory, Code 662, NASA/Goodard 
              Space Flight Center, Greenbelt, MD 20771
         \and
              Department of Physics and Astronomy, Johns Hopkins University, 
              Baltimore, MD 21218
             }

   \date{Received ...; accepted ...}

 
  \abstract
   {The comparison of observed counts in a given sky direction with 
    predictions by Galactic models yields constraints to the spatial 
    distribution and the stellar birthrate of young stellar populations.
    In this work we present the results of the analysis of the stellar content
    of the XMM-Newton Bright Serendipitous Survey (XBSS). 
    This unbiased survey includes
    a total of 58 stellar sources selected in the 0.5 -- 4.5 keV energy band,
    having a limiting sensitivity of \mbox{$10^{-2}$ cnt~s$^{-1}$} and
    covering an area of 28.10 sq. deg.}
   {Our main goal is to understand the recent star formation history of the 
    Galaxy in the vicinity of the Sun.} 
   {We compare the observations with the predictions obtained with XCOUNT,
    a model of the stellar X-ray content of the Galaxy. The model predicts 
    the number and properties of the stars to be observed in terms of 
    magnitude, colour, population and $f_\mathrm{x}/f_\mathrm{v}$ 
    ratio distributions of the 
    coronal sources detected with a given instrument and sensitivity 
    in a specific sky direction.}
    {As in other shallow surveys, we observe an excess of 
    stars not predicted by our Galaxy model. Comparing the colours of the 
    identified infrared counterparts with the model predictions, we observe
    that this excess is produced by yellow (G+K) stars.
    The study of the X-ray spectrum of each source reveals a main
    population of stars with coronal temperature stratification typical of 
    intermediate-age stars. As no assumptions have been made for the 
    selection of the sample, our results must be representative of the 
    entire Solar Neighbourhood.
    Some stars show excess circumstellar absorption 
    indicative of youth.}
   {}


   \maketitle
%

    \begin{figure*}
    \centering
    \includegraphics[width=17.0cm]{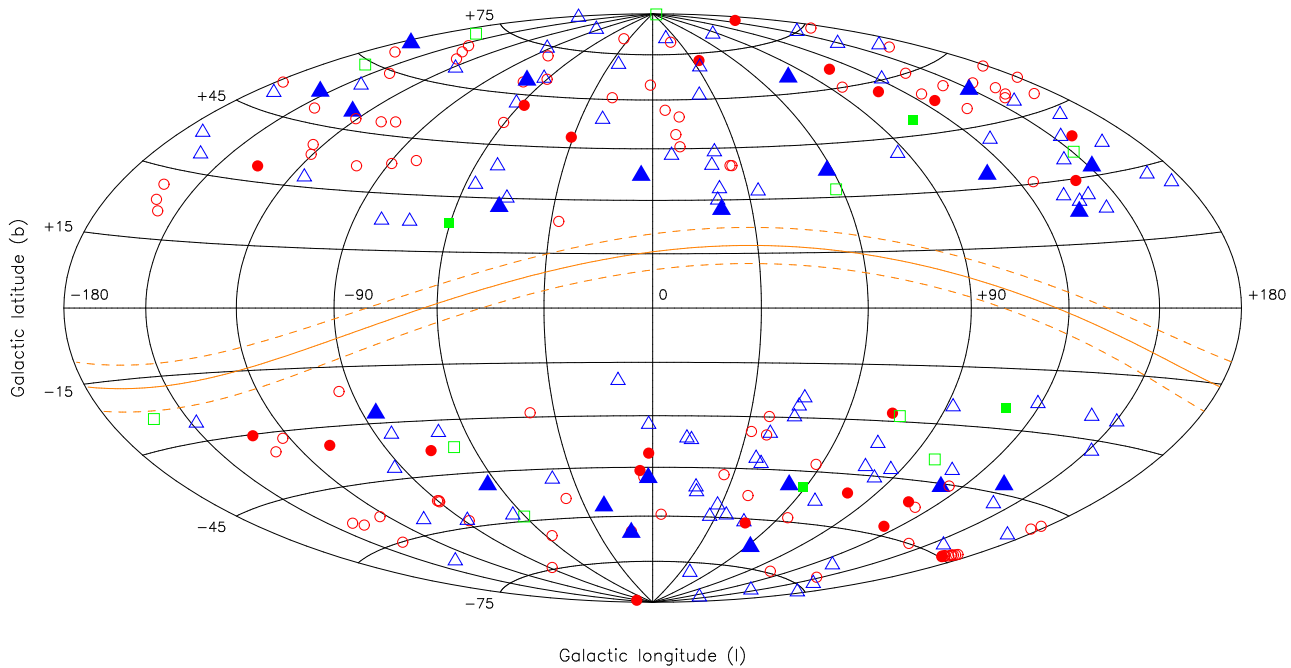}
    \caption{Galactic distribution of the fields of the XBSS \citep{cec04}:
             circles,
             triangles and squares are used for representing observations
             carried out using the Thin, Medium and Thick filter in MOS2,
             respectively. Filled symbols are plotted for those fields where
             stellar counterparts have been identified.
             The position of the Gould Belt is shown (continuous line),
             with a width of $\pm 5^\circ$ (dashed lines).}
             \label{fig:xbss}
    \end{figure*}

\section{Introduction}
\label{sec:intro}

    %
    It is well known from the study of young open clusters that magnetic
    activity traced by coronal X-ray emission evolves during the stars life.
    The average $L_{\rm x}$ decreases from \mbox{$\sim 10^{30}$ erg s$^{-1}$}
    to \mbox{$\sim 10^{27}$ erg s$^{-1}$} during the life of
    normal late-type stars \citep{mic85, mic88, gue97, fei99}. This decrease
    mostly occurs between the Zero Age Main Sequence (ZAMS) 
    and the solar age, while it is nearly negligible during the pre-main 
    sequence (PMS) phase. As a consequence, X-ray flux-limited 
    surveys detect young (PMS and ZAMS) stars up to larger distances 
    than old (main sequence, MS) ones. 
    Hence, shallow stellar X-ray selected samples will
    be dominated by young sources while old stars will be the dominant
    population in deep high-latitude stellar X-ray samples, since young sources
    have smaller scale heights than the limiting distance of the survey
    \citep[see][ and references therein]{mic03}.

    Stellar X-ray surveys play an important role in understanding the
    recent star formation history in our Galaxy in the last billion years
    \citep[eg.][]{fav92}. The comparison between X-ray observations and
    predictions derived from X-ray galactic models can constrain the
    properties of stellar populations, such as their spatial distributions
    and the stellar birthrate in the solar neighbourhood.
    Thus, X-ray surveys are a very useful tool for studying the stellar 
    population in the (near) Galaxy.
    %
    
    A computational model for the prediction and 
    analysis of the distribution of galactic X-ray stellar coronal sources
    counts, XCOUNT, was developed by \citet{fav92}. The model allows the 
    prediction of the number and properties, in terms of magnitude, 
    colour, population and  $f_\mathrm{x}/f_\mathrm{v}$ 
    ratio distributions of the
    stellar coronal sources detected with a given instrument and sensitivity
    in a specific sky direction.

    One of the first results of the comparison between predictions and
    observations of the Extended Medium Sensitivity Survey of \emph{Einstein}
    \citep[EMSS,][]{gio90} was the detection of an excess of yellow
    stars in the observations \citep{fav88} later confirmed to be due
    to young (Pleiades-age) stars \citep{fav93, sci95}. An excess of
    young stars in the solar neighbourhood was also detected by Exosat 
    and Rosat \citep[eg.][]{tag94}. These results exclude scenarios with 
    decreasing stellar birthrate in the last billion years, although they
    are unable to discriminate between a constant and an increasing stellar
    birthrate hypotheses, because of the limited sample size \citep{mic93}.
    On the contrary, the analysis of the Chandra Deep Field-North 
    \citep[CDF-N,][]{bra01} shows a lack of F, G and K dwarfs with respect 
    to the predictions \citep{fei04}. The number of dM stars agrees very 
    well with that predicted by the model but there is an excess of dF, dG 
    and dK stars in the predictions \citep{mic03, fei04}. The observed lack
    of yellow stars is in the opposite direction of the discrepancy obtained 
    with the EMSS, although it must be noticed that the CDF-N is dominated by
    old-disk stars while young (PMS and ZAMS) stars are dominant in 
    the EMSS. 
    In the same way, a first inspection of the medium-deep X-ray survey 
    HELLAS2XMM \citep{bal02} shows an excess of predicted stars 
    \citep{mic03b}.
    Old ($>$ 1 Gyr) stars dominate with the intermediate-age 
    (0.1 -- 1 Gyr) stars contributing no more than 23$\%$, 
    while the contribution of very young (0.01 -- 0.1 Gyr) stars 
    is negligible.
    %
    
    The discrepancy between the results of shallow
    and medium-deep surveys could be explained by a non-constant stellar 
    birthrate and/or incorrect assumptions on the spatial distribution 
    of stars and 
    its dependency on age in the galactic model. On one hand, an 
    increased star formation rate in the last billion years would account 
    for the excess of young stars in shallow surveys. On the other 
    hand, a density distribution decreasing with distance from the 
    Galactic plane much more steeply than the assumed exponential shape
    \citep[e.g.][]{gui96, hay97} would account for the lack of old 
    sources in medium and deep surveys. In this context, 
    Chandra and XMM-Newton surveys could be able to discriminate between
    these two possible scenarios, thanks to their high sensitivity
    \citep{mic03}.

    The XMM-Newton Bright Serendipitous Survey
    (XBSS)
    was conceived with the aim of complementing the results obtained by medium
    and deep X-ray surveys. It contains two flux limited samples of 
    serendipitous XMM-Newton sources at galactic latitudes $|b| > 20^\circ$
    (see Fig.~\ref{fig:xbss}): the XMM Bright Source Sample (BSS) containing
    the sources detected in the energy range \mbox{0.5 -- 4.5 keV} with
    count-rate \mbox{$\ge 10^{-2}$ cnt s$^{-1}$} in EPIC-MOS2, and the XMM 
    Hard Bright Source Sample (HBSS) with sources revealed in the 
    energy range \mbox{4.5 -- 7.5 keV} with count-rate $\ge 2\times10^{-3}$
    \mbox{cnt s$^{-1}$} \citep[see][for details]{cec04}.
    For a Raymond-Smith spectrum with temperature 0.7 keV, typical of 
    an active coronal source, the count-rate limit in the two 
    chosen bands corresponds to an unabsorbed 
    flux limit of $6.8 \times 10^{-14}$ \mbox{erg cm$^{-2}$ s$^{-1}$}
    and $3.7 \times 10^{-13}$ \mbox{erg cm$^{-2}$ s$^{-1}$}, respectively.
    With this sensitivity, the XBSS complements deeper XMM-Newton and Chandra
    surveys with fluxes ranging from $10^{-14}$ to $10^{-15}$ 
    \mbox{erg cm$^{-2}$ s$^{-1}$}.

    \begin{figure*}[!t]
    \centering
    \includegraphics{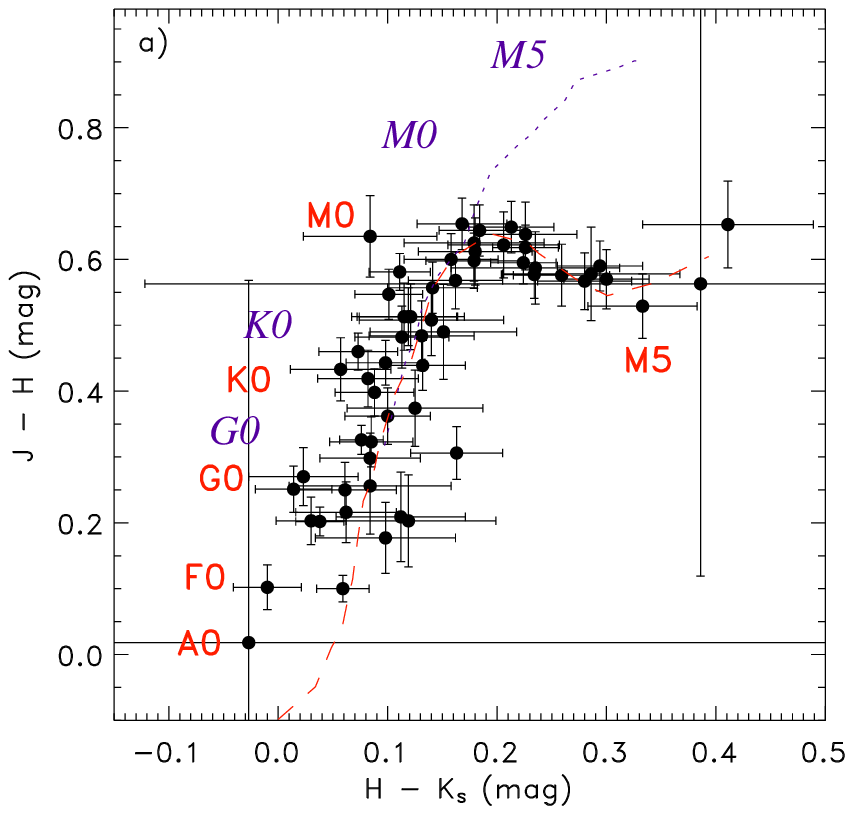}\hspace{0.3cm}
    \includegraphics{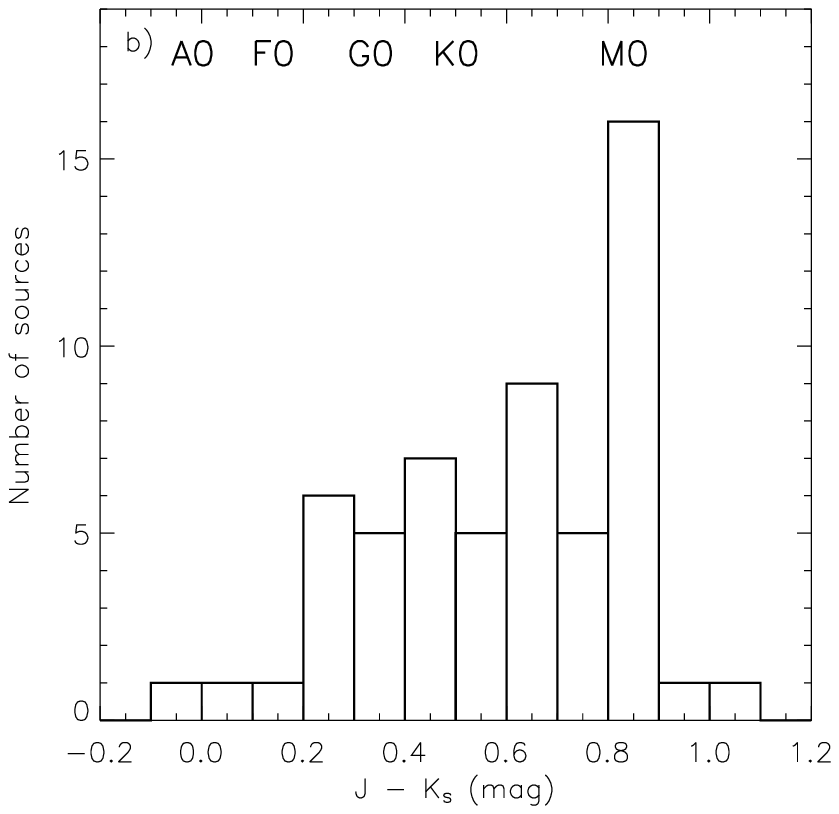}
    \caption{\textbf{a)} Colour-colour NIR diagram for the stars in
            the BSS. Dashed and dotted lines are the main
            sequence track and the giant branch
            defined by \citet{bes88} transformed here to
            the 2MASS system (see text). Spectral types of
            the main sequence are labeled with roman font, at the
            corresponding $J-H$ value but offset in $H-K_\mathrm{s}$.
            Similarly, italic font has been utilized for labeling the
            spectral types of the giant branch.
            \textbf{b)} $J-K_\mathrm{s}$ distribution of the
            sources in a).
            }
             \label{fig:jhk}
    \end{figure*}
    
    The total area covered by the survey at the flux limit is
    28.10 sq. deg. for the BSS (25.17 for the HBSS).
    The complete sample contains 400 X-ray sources,
    389 sources belonging to the BSS.
    Up to now $\sim 90\%$ of the BSS sources have been observed
    spectroscopically. 
    Fifty-eight of these sources have been identified with stars, 
    corresponding to 15\% of the BSS sample.
    In the following we will refer only to the BSS as our sample
    since all the stars belong to it while only two belong to the HBSS,
    which is largely dominated by extragalactic sources.
    %
    %
    %
    
    Thanks to the XMM large effective area, the BSS is unique since it is
    the only survey of this kind where X-ray spectroscopy can be obtained 
    for all the stars detected.
    The other particularity of the BSS is that it is unbiased at the flux
    limit; i.e., none \textit{a priori} assumption has been made when 
    selecting the sample. This makes the obtained results to be representative 
    of the entire Solar Neighbourhood.
    In addition, we should have detected all the young stars in the 
    field of view. With the sensitivity of the BSS, X-ray sources with 
    $L_\mathrm{X} \sim 10^{29}$ erg~s$^{-1}$ (typical of young stellar 
    coronal sources) could be observed at distances up to 110~pc
    (see $\S$~\ref{sec:model_desc} for a more detailed description). 
    Assuming a scale height of 100~pc for the young 
    (0.01 -- 0.1 Gyr) stellar population, 
    our sample should contain the large majority of the young stars 
    in the field of view. 

    Here we carry out an analysis of the X-ray spectra of the 58 sources
    \footnote{Fifty six of them are given in \citep{cec04}. 
    The new stars identified since then are XBS\,J040744.6-710846 and
    XBS\,J122655.1+012002.}  
    identified with stars in the BSS \citep[see][ for details on the survey 
    strategy and sample selection]{cec04} as well as a study of their
    population, e.g. their distribution in spectral type and age, 
    computed using the Galactic model XCOUNT.
    In $\S$~\ref{sec:cross}
    we determine the optical and infrared counterparts of the X-ray sources.
    In $\S$~\ref{sec:xray} we give the results of the analysis of the X-ray 
    data. A discussion on the comparison between model predictions and 
    observations is carried out in $\S$~\ref{sec:model}. Concluding 
    comments are given in $\S$~\ref{sec:conc}.


\section{Cross-identification}
\label{sec:cross}

    Infrared counterparts of our X-ray sources have been identified
    from the 2MASS\footnote{The Two Micron All Sky Survey is a joint
    project of the University of Massachusetts and the Infrared
    Processing and Analysis Center/California Institute of Technology,
    funded by the National Aeronautics and Space Administration and
    the National Science Foundation.} database. For the 
    cross-identification, a search radius of 18 arcsec has been 
    adopted bearing in mind the 15.1 arcsec XMM-Newton mirror 
    \textit{psf} half energy width, the
    \mbox{$1 - 2$ arcsec} uncertainty in the absolute position of a
    source in the EPIC detectors and the astrometric accuracy of
    \mbox{$\sim 0.2$ arcsec} of the 2MASS catalogue.
    Note that Della Ceca et al. (2004) estimate a positional error
    of the X-ray sources of the XBSS of 6 arcsec. The results of
    our cross-identification of the stellar sample with the 2MASS 
    database are in agreement with these errors.
    All the infrared counterparts are found within a radius of
    \mbox{8 arcsec} from the centroid of the X-ray source, 95\% with
    offset \mbox{$\le 4$ arcsec}. A cross-identification with the
    SIMBAD\footnote{SIMBAD Astronomical Database is operated at CDS,
    Strasbourg, France.} database and Tycho-2 catalogue has also been 
    carried out in order to identify obvious optical counterparts
    and to verify the spectral types of the infrared counterparts.

    Of the 58 stars identified to date in the BSS, 29 have an
    optical counterpart in SIMBAD, while the whole sample has been
    cross-identified with 2MASS. The infrared and optical counterparts
    are listed in Table~\ref{tab:id}, together with the $JHK_\mathrm{s}$
    colour information from 2MASS. Right ascension and declination
    of the infrared sources and positional 
    offset between the X-ray sources and their 2MASS counterparts
    are also given.
    In Fig.~\ref{fig:jhk}a we show the $J-H$ vs. $H-K$ colour-colour
    diagram for our sample. The dashed line in the figure represents
    the main sequence track as defined by \citet{bes88} adapted here
    to the 2MASS photometric system following the \mbox{Section 4} of
    the \textit{Explanatory Supplement to the 2MASS All Sky
    Survey}\footnote{The 2MASS Explanatory Supplement is available in the
    web page:\\
    http://www.ipac.caltech.edu/2mass/releases/allsky/doc/explsup.html}.
    The dotted line represents the giant branch \citep{bes88}.
    All the stars in the sample are on the main sequence track,
    which coincides with the giant branch for spectral types G0--M0.
    Nevertheless, two exotic objects
    have been optically identified (see Table~\ref{tab:id}):
    the wide binary system \mbox{WD 1631+78} (\mbox{XBS J162911.1+780442})
    and the cataclysmic variable \mbox{BL Hyi}
    (\mbox{XBS J014100.6-675328}). The former is a binary system made up of
    a white dwarf and an active M4.5 dwarf \citep[e.g.][]{coo92}
    which likely causes the observed X-ray emission \citep[see][]{sio95}. 
    The latter is a well-known polar
    -- a binary system formed by a white dwarf and a pre-main sequence
    red dwarf -- where the red spectral energy distribution of the
    companion dominates the overall colour \citep[see][]{tov01, hoa02, wat03,
    cac04}.
    
    We have used the infrared colours of the 2MASS counterparts 
    for determining the spectral type of the sources,
    which will be used in comparing observations with model predictions
    ($\S$\ref{sec:model}). For counterparts with 
    \mbox{$J-H \le 0.6$ mag.} and \mbox{$H-K_\mathrm{S} \le 0.15$ mag.} 
    we use $J-H$; for the rest of the sources we use $H-K_\mathrm{S}$. 
    The reason of such a division is that the $J-H$ index is adequate for 
    spectral classification of A to early-K stars but not for late-K and
    M dwarfs, where the differences in $J-H$ are very small. For them, the 
    $H-K_\mathrm{S}$ index is more accurate (see Fig.~\ref{fig:jhk}a). 
    Possible reddening has not been taken into account. However, no
    reddened sources are identified when comparing infrared colours with 
    the optical ones from Tycho-2, for the sources in common.
    Fig.~\ref{fig:jhk}b shows the $J-K_\mathrm{S}$ distribution of the 
    infrared counterparts.
    Sixty four per cent of the sources (including both binary
    systems containing a white dwarf companion) are K--M stars; the
    remaining are yellow stars -- spectral types F--G -- with the
    exception of \mbox{XBS J224833.3-511900}, with colours typical of a
    B dwarf but with very large errors, \mbox{XBS J080309.8+650807}, and
    \mbox{XBS J051617.1+794408} for which we obtain two possible 
    counterparts with offsets of 2.3 and 7.1 arcsec, respectively.
    The photometry of the closest counterpart is very uncertain 
    (\mbox{flag = AEE} in the 2MASS) and no conclusions about its 
    spectral type can be inferred from it. Moreover, the 2MASS detection 
    could be doubtful because of the proximity of the other source. 
    Thus, for our study we have chosen the counterpart at 7.1 arcsec.
    
    For 23 of our sources we have optical spectra from the
    optical identification campaigns. Spectral types determined using 
    this information are, in most cases, well in agreement with those 
    inferred from the infrared colours.
    Note that typical errors in the colours of these sources are 0.04\,mag,
    typically equivalent to
    2$-$3 spectral subtypes.
    %
    In Table~\ref{tab:id} we give the results of the spectral classification
    using the information of both the infrared colours and the optical 
    spectra.
    Based on infrared colours, only two sources seem to be significantly 
    reddened: XBS\,J095955.2+251549 
    and XBS\,J122942.3+015525. The difference between the infrared and 
    the optical inferred spectral types in both sources is higher than  
    acceptable for their infrared colour errors.
    Another source (XBS\,J074359.7+744057) shows discrepancy in the 
    spectral classification. In this case, the infrared colour is 
    more typical of a late-G or early-K star while the optical 
    spectrum is clearly from a late-K dwarf.
    Likewise, the two stellar systems containing a white dwarf are clearly
    identified, since the low mass companion dominate in the infrared range
    while the white dwarf is dominant in the optical spectrum.


\section{X-ray properties of the sample}
\label{sec:xray}

    \begin{figure*}
    \centering
    \includegraphics[width=18.0cm]{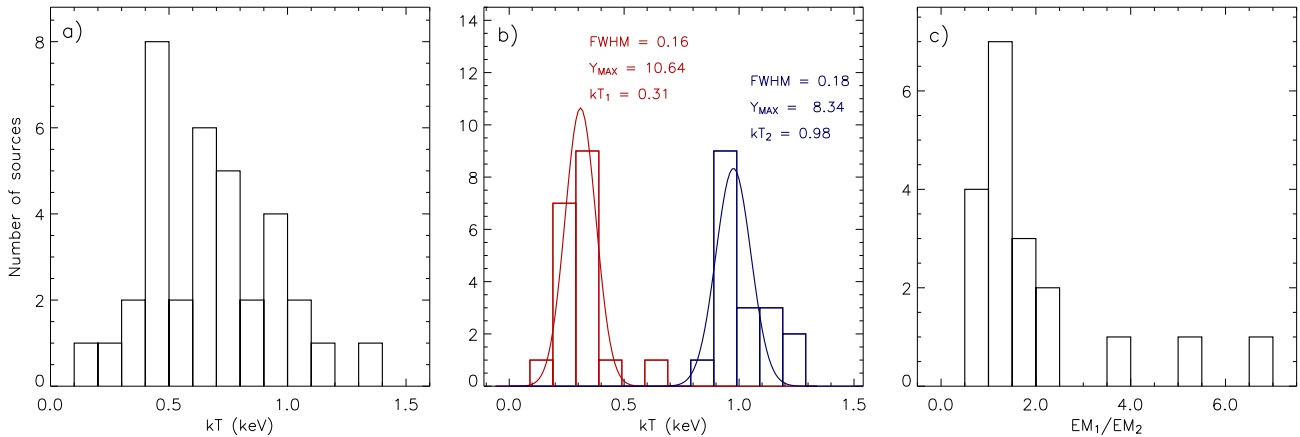}
    \caption{\textbf{a)} Temperature distribution for the sources where
             a 1T-model has been fitted. 
             \mbox{\textbf{b)} Distribution} of $T_1$ and $T_2$ of the 
             stars where two thermal components have been fitted.
             \mbox{\textbf{c)} Distribution} of $EM_1/EM_2$ of the stars 
             in figure b).
             }
             \label{fig:hist}
    \end{figure*}

\subsection{X-ray spectra and temperatures}

    We have carried out a spectral analysis of all the 58 X-ray 
    stellar sources 
    using data from the EPIC camera and the XSPEC spectral fitting 
    package \citep{arn96, arn04}. When possible, data from the PN, 
    MOS1 and MOS2 detectors have been used. In some cases, no 
    image from EPIC-PN has been collected due to both the source 
    position in the field of view and the camera
    configuration -- note that the source sample has been 
    defined using the data from the EPIC-MOS2 detector only 
    \citep[see][ for details]{cec04} without restrictions in the 
    EPIC-PN configuration -- and only MOS data have been used.

    For the analysis, we have adopted the APEC database, 
    which contains the relevant atomic data for both continuum 
    and line emission \citep{smi01} included in XSPEC software.
    Interstellar absorption has been taken into account using
    the interstellar photo-electric absorption cross-sections of
    \citet{mor83} also available in XSPEC.
    A two-temperature model has been used for those stars where a 
    1T-model does not account for the hard tail of the spectrum,  
    i.e. the least-squares solution to the 1T-model 
    gives unsatisfactory $\chi^{2}$ goodness-of-fit values. 
    Free parameters of the model are the $N_\mathrm{H}$, 
    the global abundance scaled on solar values from
    \citet{and89} and the temperature and the emission measure 
    ($EM$) for each thermal component.
    For the stars \mbox{HD 32558} and \mbox{CD-39 7717B},
    a 3T-model has been used to obtain an accurate fitting to the
    hard tail of the X-ray spectrum.

    The results of the fits are shown in Table~\ref{tab:fit}: for each
    star we give the value of the coronal temperature(s), 
    the emission
    measures rate ($EM_1/EM_2$) when two thermal components are fitted,
    the global abundance ($Z/Z_\odot$), the interstellar absorption
    ($N_\mathrm{H}$), the reduced $\chi^2$ value and associated
    degrees of freedom of the fit, the total net counts in the
    EPIC (MOS + PN) chips
    and the unabsorbed flux in the energy range \mbox{0.5$-$10 keV}.
    The error ranges refer to the $90\%$
    confidence region, or $2.706~\sigma$.
    %
    %
    For those sources with known parallax, the X-ray luminosity 
    ($\log{L_\mathrm{X}}$) has been determined (see Table~\ref{tab:selected}).
    The results show a sample dominated by active stars --\,even though 
    it is highly biased\,-- 
    with $\log{L_\mathrm{X}}$ values typical of those observed in the Hyades
    and the Pleiades. Furthermore, two of the sources (J001002.4+110831 and 
    J123600.7-395217, see Table~\ref{tab:id}) have been identified 
    with post-T Tauri stars.
    
    In Fig.~\ref{fig:hist} we plot the results in terms of temperature 
    and $EM$ distributions. For the sources where a 1T-model is adequate 
    for reproducing the spectra, 
    a broad spread of temperature is present (see Fig.~\ref{fig:hist}a).
    For them, no definite conclusions can be reached. Some of the
    sources showing a low coronal temperature have a very low $S/N$, and
    little can be said about their temperature stratification
    but that the X-ray spectra is dominated by this temperature. Several 
    stars with a mean coronal temperature $kT$ in the range $[0.5, 0.8]$ 
    keV show high $S/N$ suggesting that only one thermal component 
    is present in the corona. However, this is common only in F-type stars. 
    %
    In addition, a considerable fraction of the sources (25$\%$) 
    have a mean coronal temperature over 0.8~keV. This suggests the presence 
    of two unresolved thermal components in their X-ray spectra, the hotter 
    component contributing more than the cooler one. 
    %
    On the other hand, the spread in the temperature 
    distributions of both thermal components of the sources where a 
    2T-model has been used is remarkably lower (see Fig.~\ref{fig:hist}b). 
    A Gaussian curve has been fitted to both peaks in the temperature 
    distribution
    obtaining a $FWHM$ of 0.15 and 0.19~keV respectively, which is compatible 
    with the estimated errors (see Table~\ref{tab:fit}). These results 
    indicate that the corona of these sources can be described by two 
    thermal components, a first one peaking at 0.31~keV and a second one 
    at 0.98~keV.
    Similar results were found by a number of authors
    \citep{vai83, sch84, maj86, sch90} for late-type stars observed with 
    the \emph{Einstein} Imaging Proportional Counter \citep[IPC;][]{gor81}.
    In particular, \citet{sch90} found that a 2T-model with $kT_1 \sim 0.22$
    keV and $kT_2 \sim 1.37$\,keV is more adequate 
    than the classical 1T-model, for describing the observed X-ray spectrum 
    of most of the dwarfs in an extensive sample of late-type stars.  
    Although 
    \citet{sch84} and \citet{maj86}
    found
    that two such temperature components correspond to regions of the 
    IPC's peak sensitivity,
    \citet{sch90}
    conclude that the dichotomy is not due
    to an artifact of their analysis but musts represent an intrinsic
    property of the coronal spectra of moderately active late K and M dwarfs.
    Confirming these results, \citet{bri03} show --~using XMM-Newton
    observations~-- that the temperature 
    structure in the quasi-steady coronae of solar-like Pleiads can be 
    reproduced by two thermal components with approximately equal emission 
    measures. 

    \begin{figure*}
    \centering
    \includegraphics{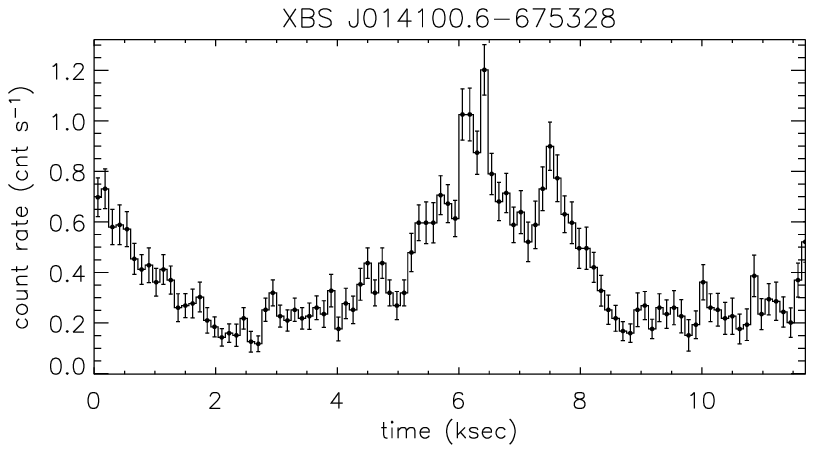}\hspace{0.3cm}
    \includegraphics{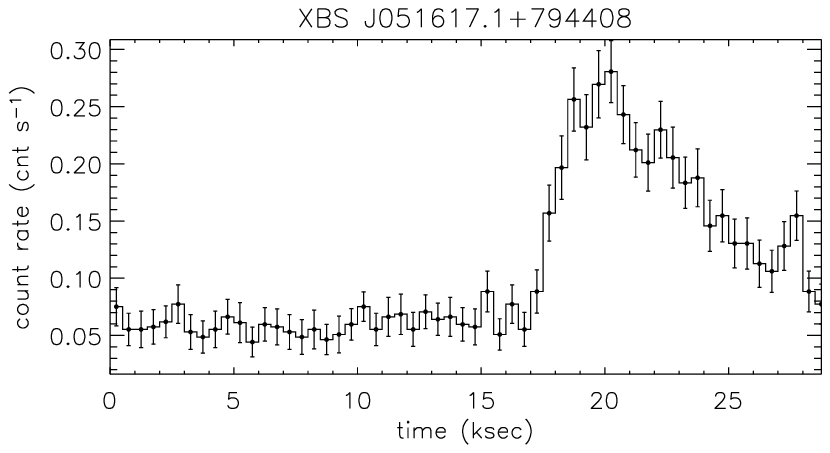}\vspace{0.3cm}
    \includegraphics{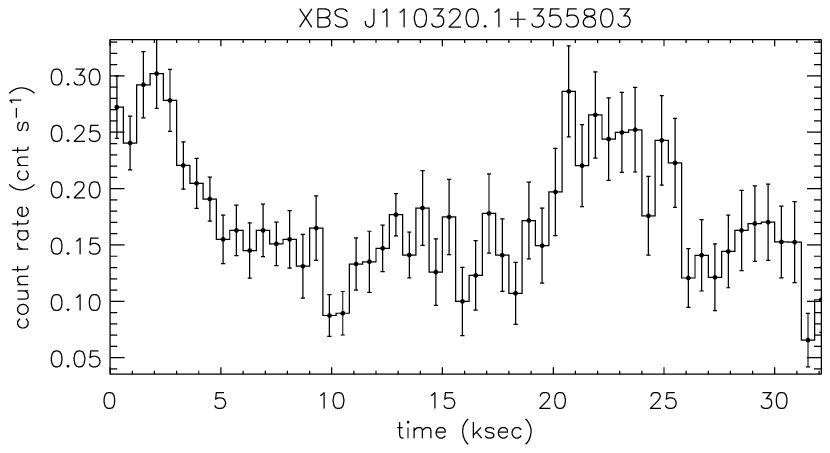}\hspace{0.3cm}
    \includegraphics{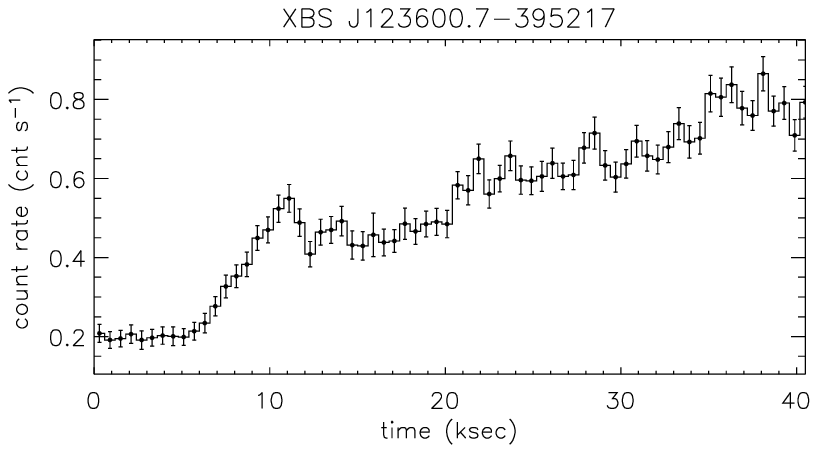}\vspace{0.3cm}
    \includegraphics{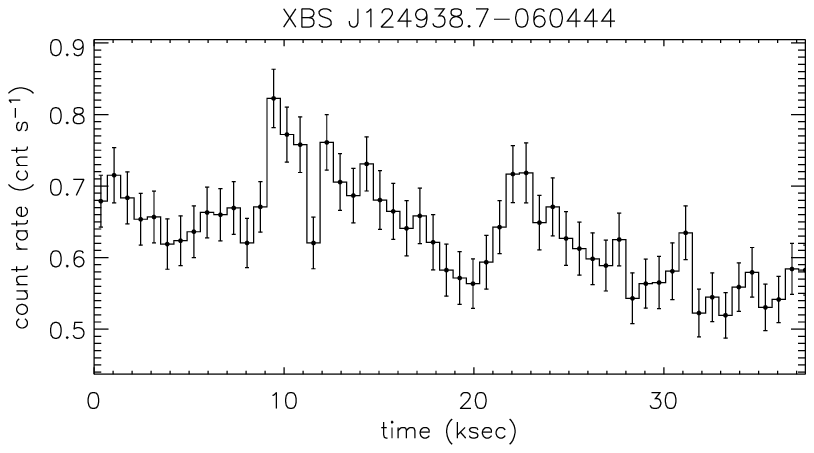}
    \caption{X-ray light curves of the most prominent variable 
             sources in our sample ($E = 0.5-10$ keV). 
             The curves have been corrected of background 
             (see text for details), GTI and dead-time effects.
             EPIC-PN data have been used except for 
             \mbox{XBS J014100.6-675328} (BL Hyi), where we have 
             chosen EPIC-MOS1 data to show also the first emission 
             decay. 
             }
             \label{fig:curves}
    \end{figure*}

    The values of the two temperatures we have found for the 
    stars in which a 2T-model is needed are compatible with the ones
    obtained by \citet{bri03} in non-saturated K stars in the Pleiades
    (see their Fig.~5). The contribution of both components of our sources
    to the X-ray spectrum is also in agreement with their results for K stars.
    In Fig.~\ref{fig:hist}c we plot the distribution of the ratio of the 
    emission measures ($EM_1/EM_2$) of the sources.
    The data show a peak around 1.0--1.5, with a spread that 
    accounts for the different contributions of the two thermal components 
    in each coronal source. Only four stars present higher contribution 
    of the hotter thermal component, all them containing more than 3000
    counts. On the other hand, three sources show $EM_1/EM_2$ values
    higher than 3, i.e. their spectrum is clearly dominated by the cooler
    component. All these stars contain less than 1000 counts. 
    Although one is tempted to look for a correlation between the number 
    of counts and $EM_1/EM_2$ -- since the higher is the coronal activity 
    the hotter is the corona of the source and the higher its count 
    rate -- no clear conclusions can be inferred from 
    these results since some of the sources with high $S/N$ 
    do not show higher contribution of the hotter component.
   
    The values of $N_\mathrm{H}$ determined from the fitting
    of the X-ray spectra are typical of nearby stars for the majority 
    of our sources. However, nine of them (15$\%$ of our sample) show
    a $N_\mathrm{H}$ significantly higher than the Galactic column 
    density. The high $N_\mathrm{H}$ derived implies that these stars
    might be surrounded by \textbf{diffuse interstellar material 
    left-over from its formation},
    since all our stars are nearby (see $\S$~\ref{sec:intro}).
    The total number of counts of some of these sources is very low and 
    no conclusions can be inferred from the fitting. In these cases, 
    the high value of $N_\mathrm{H}$ could be an artifact of the 
    fitting procedure and other solutions with similar goodness-of-fit
    values, but with different fitting parameters, could be obtained. 
    Instead, at least two of the sources 
    (XBS\,J074359.7+744057 and XBS\,J122942.3+015525) with high 
    $N_\mathrm{H}$ ($0.76\times10^{21}$ and $2.73\times10^{21}$ cm$^{-2}$, 
    respectively)
    have also a high number of counts ($\sim 700$ counts) 
    and show evidences of local reddening in their infrared colours, 
    when comparing them with the spectral type deduced from their
    optical spectrum (see Table~\ref{tab:id}). 
    The source XBS\,J095955.2+251549, with $N_\mathrm{H} = 0.76\times10^{21}$ 
    cm$^{-2}$, also shows a discrepancy between the spectral type deduced 
    from the infrared colours and that determined from the optical spectrum.
    A $N_\mathrm{H}$ of $0.6 - 2.7\times10^{21}$ cm$^{-2}$ 
    (\textbf{corresponding to $A_V \approx 0.3 - 1.3$ mag., using the 
    $(N_\mathrm{h}/A_V)_\mathrm{gal}$ relation in \citet{ryt96} and 
    \citet{vuo03}}) is compatible with that found 
    in some young stars in the nearby stellar associations TW\,Hya, 
    Tucana-Horologium and $\beta$\,Pic \citep[e.g.][]{kas03, zuc04}, where
    the residual circumstellar material is expected to dominate the 
    X-ray absorption.
    Thus, at least for these three sources, the high $N_\mathrm{H}$ seems
    not to be an artifact of the fitting process but the result of 
    X-ray absorption by circumstellar material, indicative of
    a young star yet close to its formation site. In particular, 
    they could be weak-lined T Tauri stars (WTTS) that have already 
    lost their disk since, since they do not present any IR excess in 
    the colour-colour diagram (Fig~\ref{fig:jhk}a).

\subsection{X-ray variability}
\label{sec:var}

    The light curve of the sources in our sample has been investigated
    with the aim of finding variations in their X-ray emission.
    A best extraction region has been determined for each source 
    taking their \textit{psf} into account in order to contain the 
    maximum number of photons coming from the source and the 
    minimum from the background. Besides, a background light curve
    has been extracted for background subtraction, from a region 
    of the field of view close to the source and free from other 
    contaminating sources.
    Error bars have been determined assuming Poissonian distribution of
    the photons collected by the chip in both source and background regions.
    %
    
    For our study we have utilized a Kolmogorov-Smirnov test.
    We have chosen a 
    (significance) probability $P > 0.05$ for discarding the source 
    as variable, which means we can reject the hypothesis that the 
    source is constant -- i.e. the X-ray photons 
    came from the source in a constant rate -- at the 95$\%$
    confidence level.
    Of the 58 sources composing the BSS, 39 fulfill this condition
    and other 3 are in the range 90--95$\%$. This means that 
    67$\%$ of the sources of the sample are variables. 
    Fifteen of 
    these sources show at least one flare during the observations, 
    corresponding to 26$\%$ of the sample.
    
    %
    The observed curves show very different patterns related with 
    distinct processes (see Fig.~\ref{fig:curves}). 
    For instance, \mbox{XBS J051617.1+794408} 
    and \mbox{XBS J110320.1+355803} present solar-like flare events.
    %
    These sources have been identified 
    (see $\S$~\ref{sec:cross}) with the stars \mbox{HD 32558} and 
    \mbox{HD 95735}, respectively. The latter is a well-known flare M2 dwarf. 
    On the contrary, little is known about HD 32558. SIMBAD database 
    classifies it as F8, and photometric data from Tycho-2 confirm it is an
    F star. In a study of the X-ray variability of Pleiades 
    late-type stars, \citet{mar03} show that F--G Pleiades stars are 
    variable on medium or long time scales \textbf{(days--months)}. 
    Short time (flare) variability \textbf{(i.e. time scales of hours)} 
    is also 
    observed, although it is more common in dK3--dM stars. Comparing the 
    time X-ray distribution functions with those of field stars, the authors
    demonstrate that short time variability is more common in Pleiades 
    F--G stars than in field dwarfs of similar spectral types.
    Thus, the observation of this X-ray flare suggests it is a  
    young star, although we cannot discard the existence of a cooler 
    companion producing the event.
    Other stars show light curves that may be produced by quite different
    physical processes. Thus, the continuous enhancement observed in the 
    light curve of \mbox{XBS J123600.7-395217} 
    could be
    related with accretion processes in 
    T-Tauri class objects. The source has been identified 
    (see $\S$~\ref{sec:cross}) with the 
    very young system CD-39 7717 (TWA-11), made up of a T-Tauri star 
    with spectral type A0 and an M2.5 young star and member of the
    TW Hya association (TWA). The A0 star contains
    a debris system which is believed to have a terrestrial planetary
    system \citep{low05}. The M companion is a PMS. The resolution of 
    the X-ray observation does not permit to separate the components
    and not much can be said about which star the X-ray emission 
    comes from or whether it is the sum of the emission of both. 
    Nevertheless, the A0 component is known to be under-luminous 
    in X-rays compared to the average observed X-ray emission for the 
    rest of the TWA members \citep[e.g.][]{low05} and so the observed 
    emission is likely to came from the secondary.
    %
    We have used a Chandra ACIS-S observation from the archive in which 
    the system has been serendipitously observed (Obs. ID: 2150) in order 
    to resolve both components. The position of the X-ray source in ACIS-S 
    confirms that the emission comes from the M star. 
    This agrees with the results of the cross-identification 
    (see $\S$~\ref{sec:cross}) where the source is identified with the M star.

    The other two examples of peculiar X-ray variability are 
    \mbox{XBS J124938.7-060444}, identified with the chromospheric
    active binary IM Vir, which X-ray emission shows a clear modulation
    with time, and \mbox{XBS J014100.6-675328} (BL Hyi) which show an 
    energetic flaring event decaying in one hour. This latter is a 
    well-known polar \citep[e.g.][]{tov01} observed in an 
    accretion state. 

                                                                                
\section{Model predictions}
\label{sec:model}

%
\begin{table*}[!t]
\caption{XCOUNT predicted X-ray counts using different stellar
         birthrates, compared with the observations.}
\label{tab:model_obs}
\small
\begin{center}
\begin{tabular}{l l c c c c c c c}
\hline\hline
                  & B & A & F0-F6 & F7-G & K  & M0-M5 & Total & Range$^\dag$\\
\hline
Observed$^\ddag$   & 1 & 2 &  9    &  9   & 21 & 14    & 56 \\
Constant birthrate ($\tau = \infty$)            & - & 2 & 3 & 5 & 7 & 14 & 31 & 19 -- 43\\
Slowly decreasing birthrate ($\tau =  15$ Gyr)  & - & 2 & 3 & 4 & 6 & 11 & 26 & 15 -- 35\\
Rapidly decreasing birthrate ($\tau =   5$ Gyr) & - & 2 & 3 & 3 & 5 &  7 & 20 & 12 -- 28\\
Slowly increasing birthrate ($\tau = -15$ Gyr)  & - & 2 & 3 & 6 & 8 & 18 & 37 & 24 -- 51\\
\hline
\end{tabular}
\end{center}
{\small
$^\dag$ Range of total number of stars predicted by the model, for 
        each stellar birthrate. This value has been estimated taking the relative
        error on the predicted source counts and the statistical uncertainties 
        in the model simulation into account.\\
$^\ddag$ The two binary systems containing a white dwarf
        (see $\S$~\ref{sec:cross}) have been discarded for
        the comparison.
}
\end{table*}

\subsection{Model description and assumptions}
\label{sec:model_desc}

    The X-ray Galactic model XCOUNT \citep{fav92, mic93} has been 
    used for predicting the number and properties of the coronal 
    sources expected in the BSS. The model
    is based on the stellar spatial 
    densities of the Bahcall \& Soneira model \citep{bac80, bac86}.
    Although the Bahcall \& Soneira model includes both an
    exponential disk and a spheroid, XCOUNT considers only the disk
    component due to the low X-ray luminosity of spheroid
    stars \citep[see][]{fav92}.
    With the sensitivity of the BSS, active coronal sources with
    $L_\mathrm{X} \sim 10^{29}$ \mbox{erg s$^{-1}$} could be observed
    at distances up to $\sim 110$ pc, while the maximum distance at which
    a typical quiet star with $L_\mathrm{X} \sim 10^{27}$ \mbox{erg s$^{-1}$}
    might be visible is $\sim 10$ pc.
    Then, the approximation to a disk is 
    quite realistic since we are observing only stars in our neighbourhood.
    Note that the scale height adopted by \citet{bac86} is $350 \pm 50$ pc
    for main sequence stars and $250 \pm 100$ for giants.
    On the other hand our model is not simply a disk. The spatial 
    distributions of the disk population in the original optical model 
    has been modified to separate several distinct 
    subpopulations of A--M stars of different ages. 
    This introduces an explicit dependence of the stellar spatial 
    distribution and density on stellar age \citep{mic93}.
  
    Different scale heights have been adopted for distinct
    age ranges: 100, 200 and 400 pc for \mbox{0.01--0.1}, \mbox{0.1--1}, 
    and $> 1$ Gyr respectively. They have been 
    determined taking the results of \citet{bas92} into account. 
    Assuming the Galaxy to be formed by 
    a disk isothermal in the z-direction \citep{wie83} the scale height 
    ($H$) is correlated with the velocity dispersion ($\sigma_\mathrm{z}$) 
    as $H \propto \sigma^2_\mathrm{z}$. At the same time, 
    the velocity dispersion is proportional to $t^\mathrm{n}$ 
    \citep{vil85} and so the scale height is proportional to the age 
    as $H \propto t^\mathrm{2n}$. The value of $n$ must be determined 
    from empirical data. In particular, \citet{bas92} obtain 
    $2n = \frac{2}{3}$ using data given in \citet{rob86}. 

    Three different X-ray 
    luminosity functions (XLFs) have been utilized for the distinct 
    age ranges: the youngest stars with ages \mbox{0.01--0.1} Gyr are 
    assigned the ROSAT XLF of the Pleiades members \citep{mic96}; 
    intermediate-age stars with ages \mbox{0.1--1} Gyr are assigned 
    the ROSAT XLF of the Hyades \citep{ste95}; and the older ones 
    ($t > 1$ Gyr) are assigned the XLF
    of \emph{Einstein} studies of nearby old-disk stars 
    \citep{sch85, mag87, bar93}.
    The XLFs have been appropriately converted to the bandpass of the 
    observations, assuming a given input spectrum 
    \citep[see][ for details]{fav92}.
    Such a set of XLFs permits to 
    evaluate the contribution to the total number of stars of different 
    age populations without losing information about the assumed 
    optical Galaxy model \citep{fav92}. 

    In general, predicted optical star counts are not strongly influenced
    by age distribution of the disk populations \citep{mic93} because
    both optical luminosity and colour do not change substantially  
    with stellar evolution during the main sequence lifetime. On the 
    contrary, due to the high dependence of the X-ray luminosity with
    age \citep{mic85, mic88, gue97, fei99} the assumed age distribution
    has significant influence on the predicted X-ray source counts.
    Here we have performed model computations with the three age subgroups 
    having relative density on the plane derived from an exponential law:

    \begin{equation}
      \Psi = A \ e^{-t/\tau}
    \end{equation}
  
    where $\tau$ can take the values $\infty$, 15, 5, and $-15$ Gyr,
    corresponding to constant, slowly decreasing, rapidly decreasing, and
    slowly increasing stellar birthrate respectively, according to
    the results by \citet{bas92}.

\subsection{XCOUNT model results}

    \begin{figure}
    \centering
    \includegraphics[width=4.3cm]{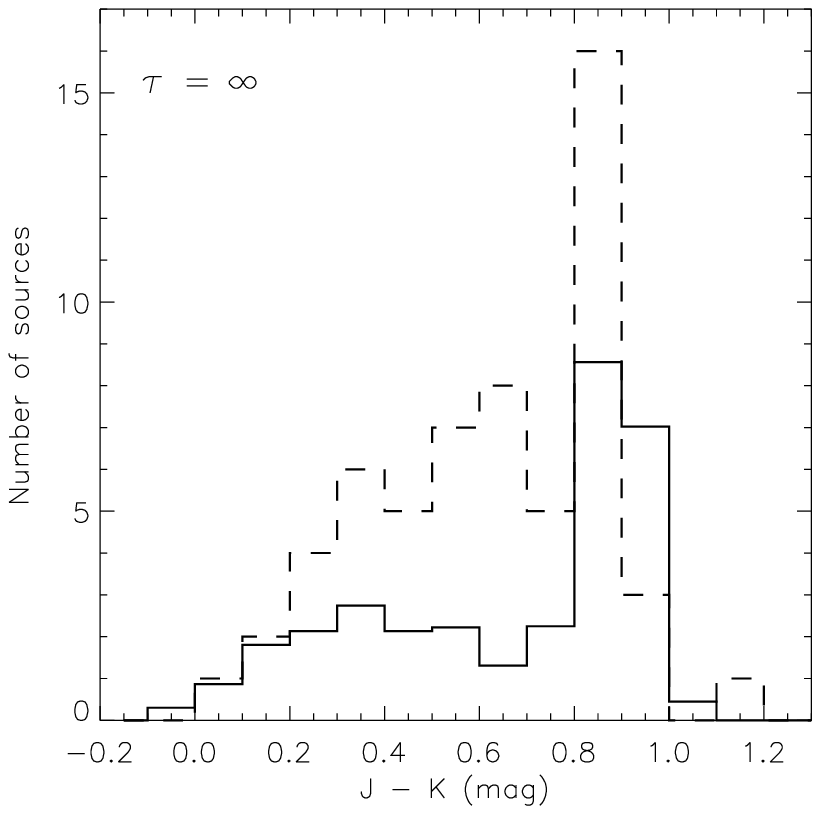}
    \includegraphics[width=4.3cm]{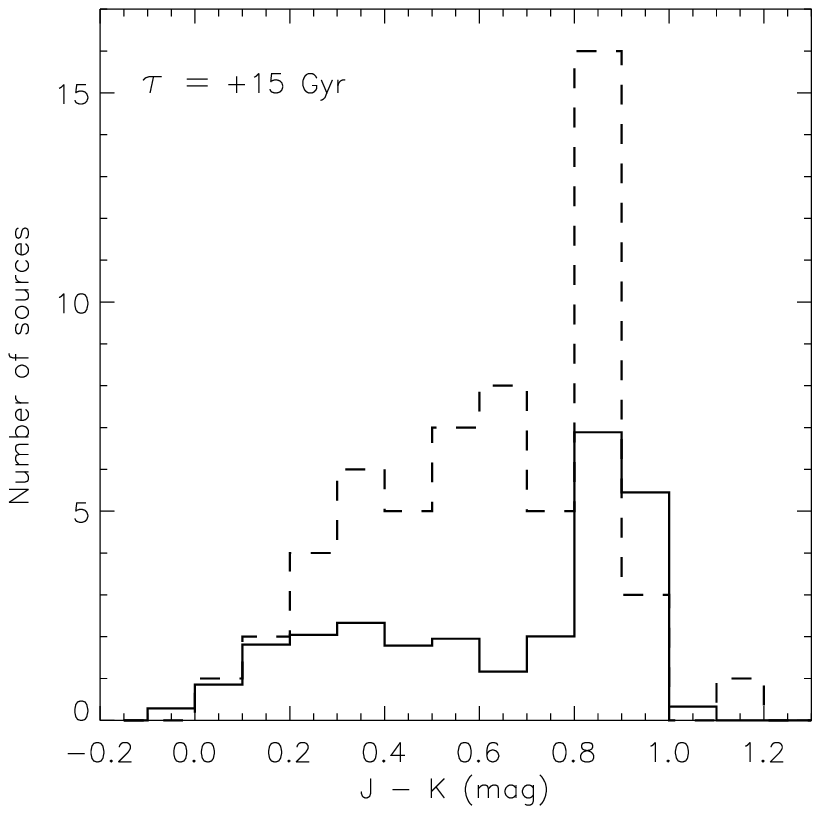}
    \includegraphics[width=4.3cm]{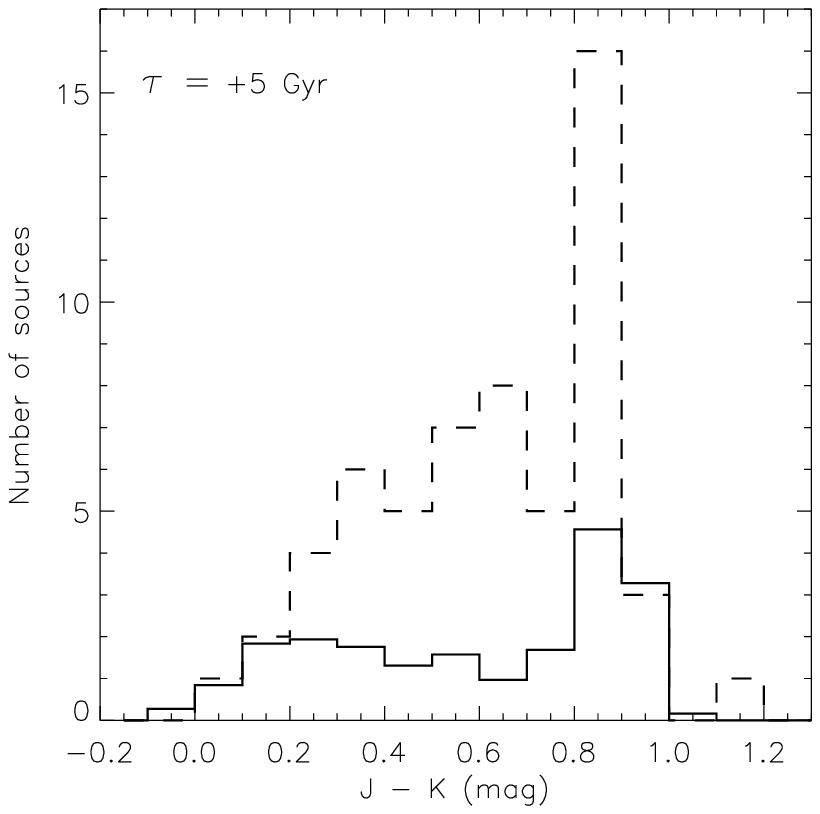}
    \includegraphics[width=4.3cm]{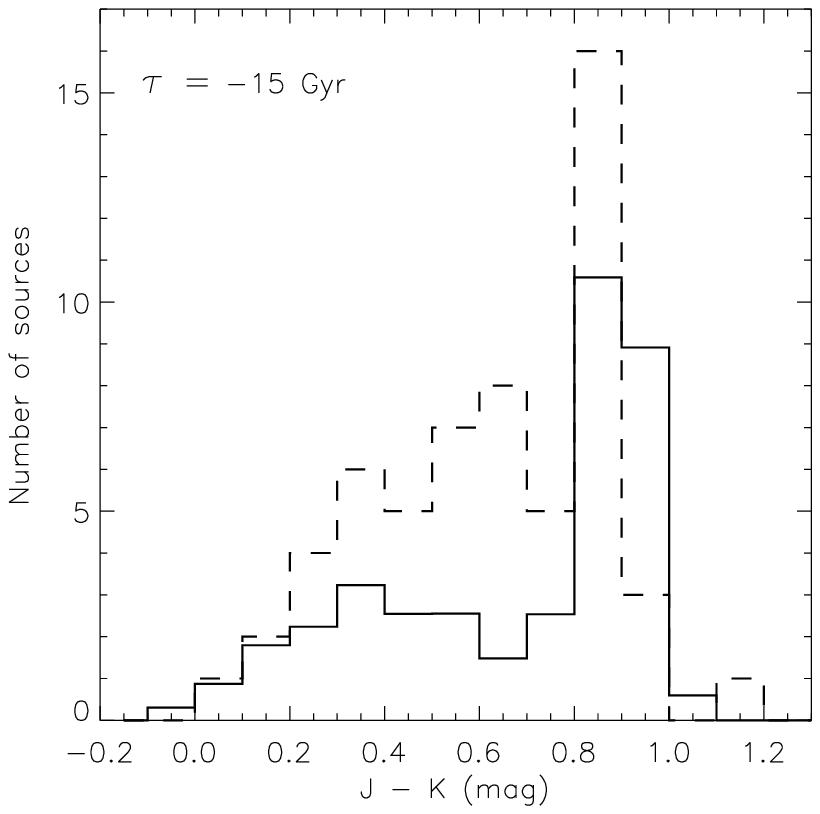}
    \caption{Colour distributions predicted by XCOUNT (continuous line)
             compared with the observed one (dashed line)
             for different stellar birthrates assuming an exponential
             law $\Psi = A e^{-t/\tau}$.
            }
             \label{fig:jk_xcount}
    \end{figure}

    \begin{figure*}[!t]
    \centering
    \includegraphics[width=8.5cm]{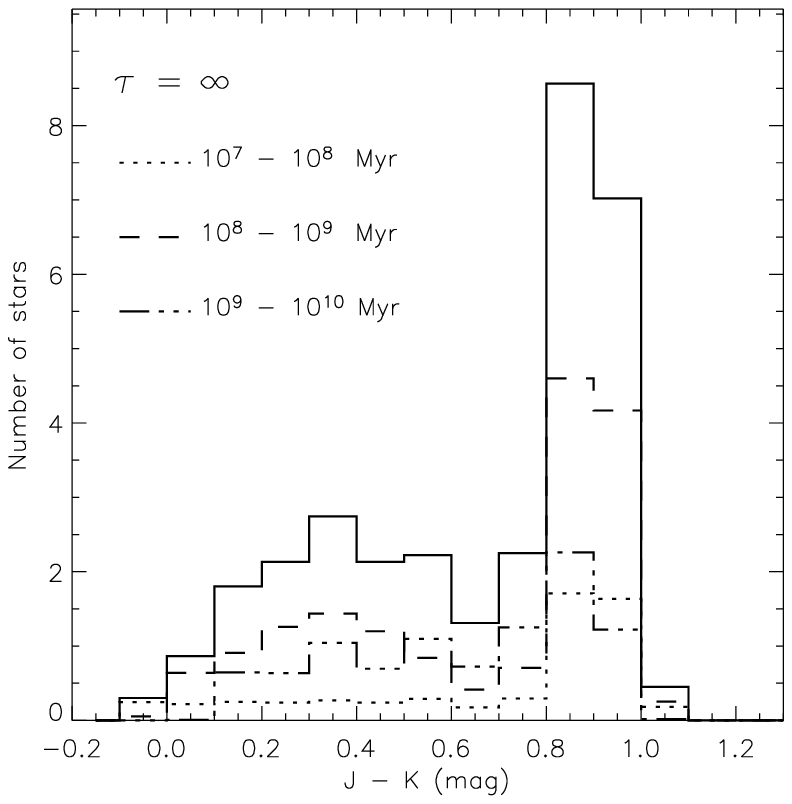}
    \includegraphics[width=8.5cm]{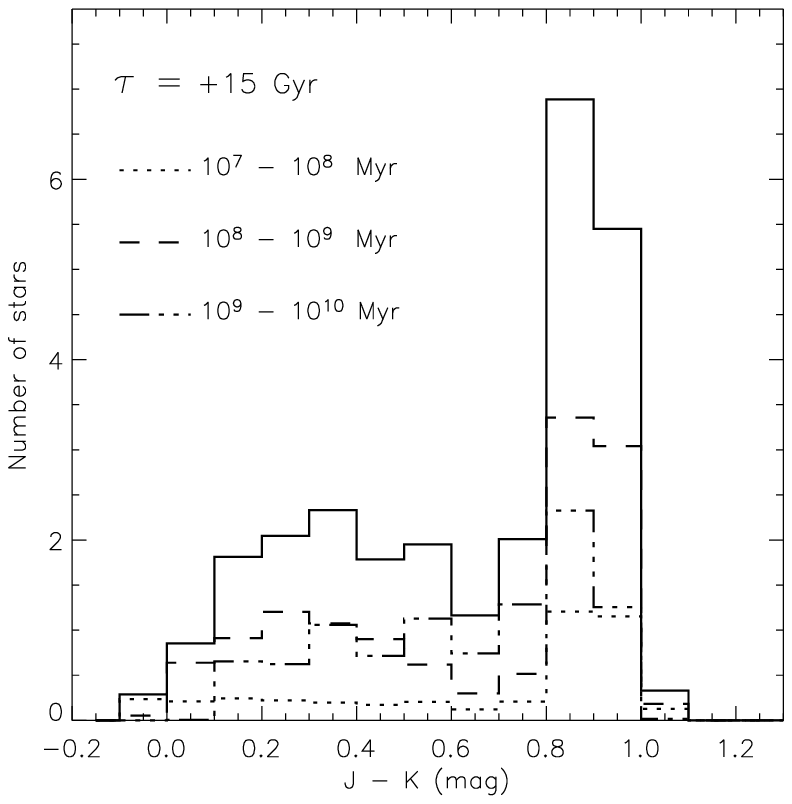}
    \includegraphics[width=8.5cm]{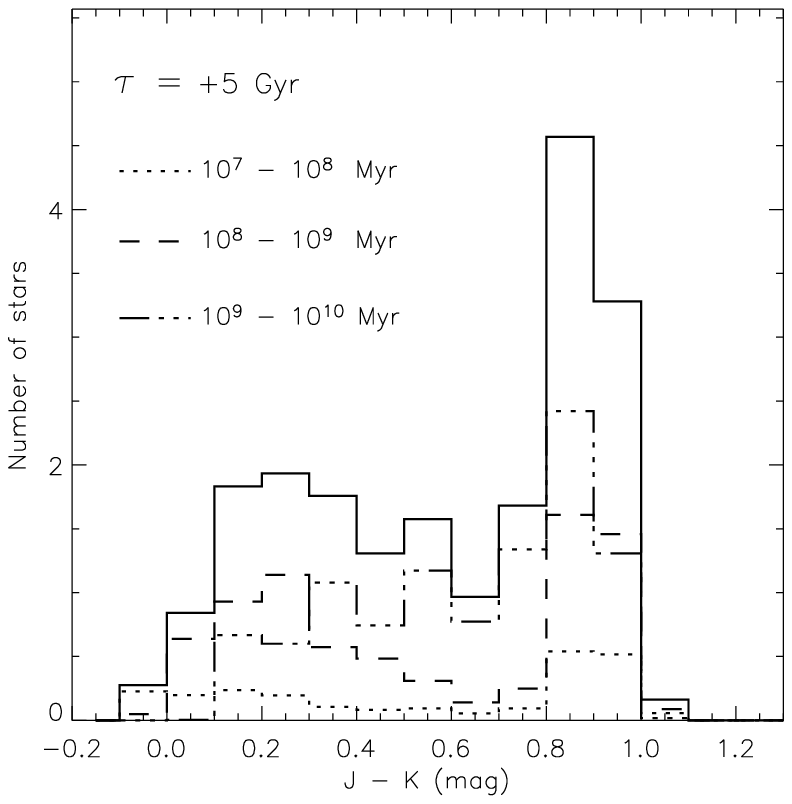}
    \includegraphics[width=8.5cm]{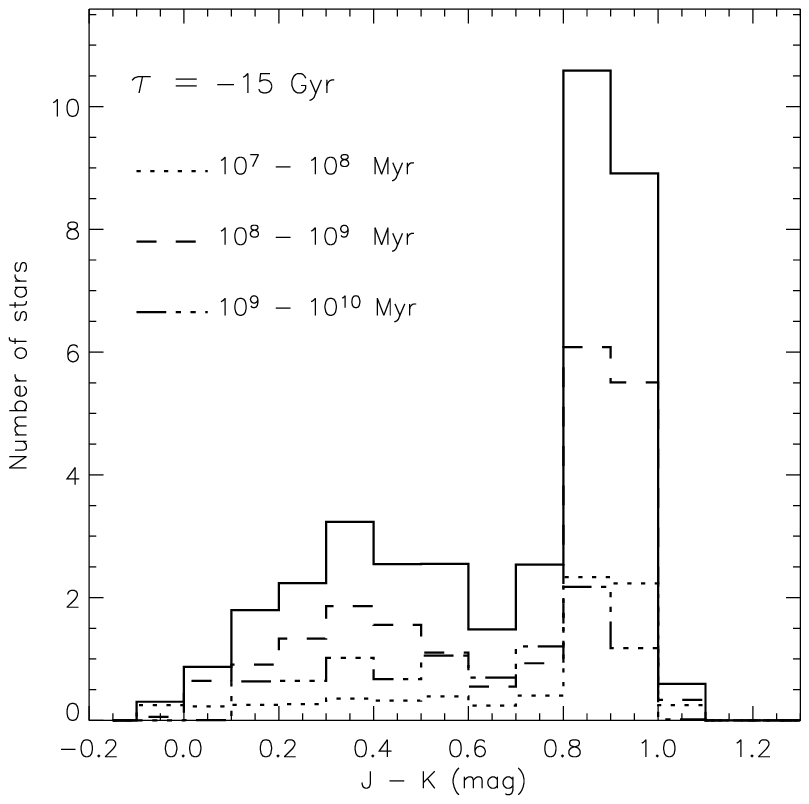}
    \caption{Colour distributions predicted by XCOUNT
             for different stellar birthrates assuming an exponential
             law $\Psi = A e^{-t/\tau}$. The contribution of different
             age-subgroups to the total source counts is shown
             in each plot.
            }
             \label{fig:xcount_sfrs}
    \end{figure*}

    In Fig.~\ref{fig:jk_xcount} we show the predictions of the model
    with relative density distributions for the three age subgroups 
    given by the above mentioned stellar birthrates. 
    The numbers are summarized in 
    Table~\ref{tab:model_obs}. The 2MASS $J - K_\mathrm{S}$ colours 
    of the observed X-ray sources infrared counterparts have been 
    converted to the Bessell \& Brett photometric system 
    -- utilized by the model --
    using the transformations given in the \textit{Explanatory Supplement 
    to the 2MASS All Sky Survey}, for comparing observations with predictions.
    
    A first inspection of the results suggests the stellar birthrate to
    be an increasing function of time since the total number of
    predicted sources is in better agreement with the observed ones
    when using $\tau = -15$ Gyr\footnote{
    Note that typical values for the relative error on the predicted source
    counts -- for the error propagation of the statistical errors associated
    with the X-ray luminosity functions -- are in the range 15-30$\%$
    \citep{fav92}. Taking this and the $\sqrt{N}$ statistical uncertainties
    in the model simulation into account \citep{fei04} we estimate a range
    of 19--43, 15--35, 12--28, and 24--51 predicted sources for stellar
    birthrates with $\tau = \infty$, +15, +5, and $-15$ Gyr respectively.}.
    However, the number of predicted M stars using this birthrate
    is quite above the observed one, even taking the error propagation in
    the model into account. On the contrary, predictions made for M stars
    using both constant or slowly decreasing birthrates are quite in agreement
    with the observations. Note as well that the slowly increasing stellar 
    formation rate (SFR) gives results very similar to those of the 
    constant birthrate. 

    On the other hand, the model does not account for the number of observed 
    FGK stars using any of the birthrates analysed here.
    An excess of yellow stars is also observed in other shallow surveys
    \citep{fav88, mic03} that cannot be reproduced by Galactic models
    using smooth shapes of the SFR. In particular, \citet{fav93} and 
    \citet{sci95} confirm the excess of yellow stars in the Extended 
    Medium Sensitivity Survey of \emph{Einstein} \citep[EMSS,][]{gio90} to be 
    due to young stars.
    Similarly, \citet{mic06} find an excess of FGK stars in the ROSAT 
    North Ecliptic Pole survey \citep[NEP,][]{hen01}.
    This result excludes scenarios with decreasing 
    stellar birthrate, but does not discard continuous SFR with 
    one or more burst. Studies of the IMF \citep{mil79} and metallicity
    distribution in F stars \citep{twa80, car85} show that the SFR 
    has basically remained constant over the age of the Galactic disk, 
    or that it might have been smaller in the past \citep{mil79}. 
    Our results show that the constant and the slowly increasing 
    birthrate hypothesis are nearly equivalent and none of them 
    reproduce the counts for yellow stars. 
    %

\subsection{The nature of the FGK excess}

    From the results presented above one can argue that no smooth 
    stellar birthrate can account for the observed distribution of stars. 
    Although the total number of sources can be quite well reproduced using
    a slowly increasing SFR, the observed and predicted distributions in 
    spectral type are different: there is an excess of stars with spectral 
    types F--K in the observations and there is a lack of observed
    M stars that compensates the excess of FGK stars.
    Taking into account that using different scale heights makes the total 
    number of stars vary but not the relative distribution in spectral types,
    and assuming that the
    stellar birthrate is indeed smooth --~no definite evidences of star-burst 
    are observed in the recent history of the disk 
    \citep[see][but see also Barry 1988]{twa80, car85}~-- 
    the discrepancy between predictions and observations must be due to 
    a stellar population not accounted for by the model.
    
    %
    In Fig.~\ref{fig:xcount_sfrs} we show the relative contribution of 
    the stars in the three age ranges to the total stellar counts, when 
    using SFR with $\tau = \infty$, +15, +5, and $-15$ Gyr, respectively.
    In the cases of rapid or slowly decreasing birthrates, the 
    X-ray population is dominated by intermediate age and old stars with 
    almost the same contribution. However, when using a constant or a 
    slowly increasing SFR
    the intermediate age sources are clearly dominant. In all the cases,
    the young stars give a small contribution (less than 20$\%$). 
    %
    The XCOUNT model allows the determination of the number of sources of 
    any type expected in an X-ray observation
    given their spatial distribution and luminosity function,
    since they are treated as external parameters \citep{fav92}. Thus, 
    we have used XCOUNT to predict the number of disk giant stars 
    observed in the BSS, using the spatial distribution of the Bahcall 
    and Soneira model, a scale height of \mbox{250 pc} and a coronal 
    temperature of \mbox{1.1 keV} \citep{sch90}. The result is 
    that no more than 1 giant is expected to be detected in the sample, 
    which discard giants as the cause of the observed excess of yellow
    stars. 
  
\begin{figure}
    \centering
    \includegraphics[width=8.5cm]{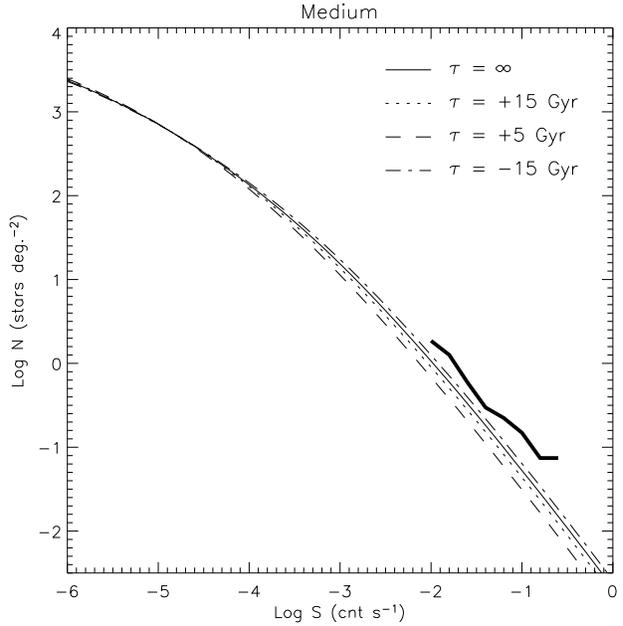}
    \caption{Log $N$--Log $S$ predicted assuming $\tau = \infty$, +15,
            +5 and -15 Gyr, compared with the BSS results (thick solid
            line) for the fields observed with the medium filter
            (energy band 0.5--4.5 keV). 
            For the comparison we have used an standard high
            latitude direction $b = 60^\circ$.
            }
             \label{fig:lnls}
    \end{figure}

    Bearing the results discussed above in mind, we infer that only
    two populations can be responsible of the excess in the BSS: active
    binaries -- \mbox{RS CVn} and \mbox{BY Dra} systems -- and/or very
    young objects. 
    In Fig.~\ref{fig:lnls} we plot the $\log N$--$\log S$ 
    predicted at high latitude for 
    the four stellar birthrates, compared with observations in the fields 
    observed with the medium filter (the results with the other filters are 
    equivalent). 
    The observed cumulative curve is clearly above the predictions.
    At low count-rates, the curve is almost parallel to the predicted 
    $\log N$--$\log S$ 
    but at high count-rates ($\ge 6 \ 10^{-2}$ s$^{-1}$) there is a bump
    that could result from the presence of a population of sources with high 
    X-ray luminosity and with low scale height. This would show a scenario 
    of recent stellar birthrate enhancement, which is in agreement 
    with the results of the EMSS \citep{fav93, sci95}.
    On the other hand, if we assume the scenario with an increasing stellar 
    birthrate to be correct --~as suggested by the better fit to the total 
    number of stars (see Table~\ref{tab:model_obs})~-- we would be observing 
    less M stars than those predicted. A possible hypothesis for the observed 
    excess of yellow stars in such a scenario with a increasing stellar 
    birthrate and a lack of red stars is the existence of a number of 
    X-ray M dwarfs in binary systems with a yellow primary. 
    %
 
\section{Discussion and conclusions}
\label{sec:conc}

    We have studied in detail the stellar content of the XMM Bright 
    Serendipitous Survey. 2MASS counterparts have been identified 
    for each one of the 58 X-ray sources. This unbiased stellar sample has 
    been characterized in the sense of coronal temperature, global abundance
    and column density, and a study of the light curves has been 
    carried out to reveal the X-ray variable population.
    Taking into account that the sample is unbiased, the results must be
    typical of the entire Solar Neighbourhood at the flux limit of the
    survey.
    This is the first time in which a complete X-ray 
    spectral characterization has been carried out for a survey of this kind, 
    thanks to the XMM-Newton effective area.
    In general, the coronal temperature stratification of our sources 
    is typical of moderately active stars. Their spectrum can be fitted 
    using a 2T-model with $kT_1 = 0.32 \pm 0.08$ keV and 
    $kT_2 = 0.98 \pm 0.09$ keV. Only for two sources, a third thermal 
    component has been necessary for fitting the hard tail of the spectrum.
    For other sources, with low S/N, a 1T-model is adequate for reproducing
    the spectrum. The value of the temperature in these cases is not 
    representative of the temperature stratification but only of the 
    dominant thermal component.
    Of the 58 sources, 42 present variability in their light curves,
    corresponding to $67\%$ of the whole sample. Fifteen of these sources
    ($26\%$ of the whole sample) show at least one flare during the 
    observations.
   
    We have performed a detailed comparison between the observed stars
    and the predictions computed using the XCOUNT model with four
    different stellar birthrates.
    From the comparison of the expected number of stars of each 
    spectral type with that inferred from the 2MASS counterparts, 
    we show that a decreasing stellar birthrate scenario 
    is not probable since the total number of predicted stars
    is clearly below the observed one, even taking model uncertainties 
    into account. A constant SFR can reproduce well the number of A and M 
    sources observed, but still underestimates the total number of
    observed stars. On the contrary, an increasing birthrate seems more 
    suitable to the BSS: the total number of predicted sources fit
    well with the observations bearing in mind model uncertainties, 
    although it overestimates the total number of M stars.
    On the other hand, the excess of FGK stars can not be reproduced using 
    only a smooth stellar birthrate. Assuming that the stellar birthrate 
    is indeed smooth in the last billion years, we infer that the discrepancy
    between observations and predictions must be produced by a stellar
    population not directly taken into account in the model.
    
    In the $\log N$--$\log S$ diagram, the cumulative curve of the BSS
    sources is above the predicted one for the four stellar birthrates
    analyzed here. 
    The BSS curve runs parallel to the predicted ones for low count-rates,
    but at higher count-rates ($\ge 6 \ 10^{-2}$ s$^{-1}$) a bump is clearly
    observed. This again suggests the existence of a population with high 
    X-ray luminosity and low scale height contributing to the sample. 
    The high percentage of flare stars observed in the sample is
    in agreement with this hypothesis.
    Nevertheless, the lack of observed M stars with respect to the predictions
    when using the increasing stellar birthrate suggests another explanation:
    the existence of a number of X-ray M stars in binary systems with a 
    yellow primary. 
    
    We have shown that a scenario with a decreasing stellar birthrate
    is not suitable for the observed number of stars and their distribution 
    in spectral type. Furthermore, the situation with a constant SFR 
    underestimates the total number of observed sources and does not seem
    to be appropriate. On the contrary, we show that a scenario with an 
    increasing stellar birthrate can account for the total number of stars, 
    and that the discrepancy in the distribution of the stars in spectral type 
    between predictions and observations could be due to a population 
    of young stars with low scale heights or to a number of M dwarfs in 
    binary systems with a yellow primary.
    %
    %

    Finally, we have shown that \textbf{in several stars of the sample, 
    some residual material left-over from its formation could be present}.
    The value of
    $N_\mathrm{H}$ determined for these stars is compatible with the
    optical absorption and column density observed in stars of the
    young stellar associations TW\,Hya, Tucana-Horologium and
    $\beta$\,Pic. This agrees with the scenario with an increasing 
    stellar birthrate and a population of young stars in the 
    Solar neighbourhood. These stars would be good targets for examining 
    the evolution of gaseous circumstellar disks around young stars.

                                                                                
\begin{acknowledgements}
J. L\'opez-Santiago acknowledges support by the Marie Curie Fellowships
contract No. MTKD-CT-2004-002769.
The XMM BSS project has received financial support by the Italian Space 
Agency (ASI), by the Ministero dell'Instruzione Nazionale, Universit\`a
e Ricerca Scientifica (MIUR) and by the italian Istituto Nazionale di 
Astrofisica (INAF). We also thank the TNG, ESO and Calar Alto Time 
Allocation Committee for a generous and continuous allocation of 
observing time during the last years. We thank Tommaso Maccacaro for the
strong support given to the XMM/Newton BSS project.
This publication makes use of data products from the Two Micron All Sky 
Survey, which is a joint project of the University of Massachusetts and 
the Infrared Processing and Analysis Center/California Institute of 
Technology, funded by the National Aeronautics and Space Administration 
and the National Science Foundation.
This research has made use of the SIMBAD database, operated at CDS, 
Strasbourg, France.
\end{acknowledgements}

\small
\begin{landscape}
\setlength{\topmargin}{5.5cm}
    \begin{longtable}{llccrrrrccc}
    \caption{\label{tab:id}
            Infrared counterparts identified for the stellar sample.
            Colour information from 2MASS is provided. Errors
            in the colours refer to the total photometric uncertainty
            as described in 2MASS. The two sources marked with $\dag$
            are the two binary systems with a white dwarf.}\\
    \hline\hline
    \noalign{\smallskip}
    Id. (XBSS $\dots$) & Optical counterpart &
    \multicolumn{6}{c}{2MASS counterpart} &  & 
    \multicolumn{2}{c}{Spectral type}\\
    \cline{3-8}\cline{10-11}
    \noalign{\smallskip}
    &  & RA            & Dec          &
    \multicolumn{1}{c}{Offset} & \multicolumn{1}{c}{J} &
    \multicolumn{1}{c}{H} & \multicolumn{1}{c}{K$_\mathrm{s}$} \\
    \noalign{\smallskip}
                       &            & (J2000.0)     & (J2000.0)    &
    \multicolumn{1}{c}{arcsec} & \multicolumn{1}{c}{(mag)} &
    \multicolumn{1}{c}{(mag)} & \multicolumn{1}{c}{(mag)} &  &
    IR count. & Opt. spectrum\\
    \noalign{\smallskip}
    \hline
    \endfirsthead
    \caption{continued.}\\
    \hline\hline
    \noalign{\smallskip}
    Id. (XBSS $\dots$) & Optical counterpart &
    \multicolumn{6}{c}{2MASS counterpart} &  &
    \multicolumn{2}{c}{Spectral type}\\
    \cline{3-8}\cline{10-11}
    \noalign{\smallskip}
    &  & RA            & Dec          &
    \multicolumn{1}{c}{Offset} & \multicolumn{1}{c}{J} &
    \multicolumn{1}{c}{H} & \multicolumn{1}{c}{K$_\mathrm{s}$} \\
    \noalign{\smallskip}
                       &            & (J2000.0)     & (J2000.0)    &
    \multicolumn{1}{c}{arcsec} & \multicolumn{1}{c}{(mag)} &
    \multicolumn{1}{c}{(mag)} & \multicolumn{1}{c}{(mag)} &  &
    IR count. & Opt. spectrum\\
    \noalign{\smallskip}
    \hline
    \endhead
    \noalign{\smallskip}
    \hline
    \noalign{\smallskip}
    \endfoot
    \noalign{\smallskip}
    J001002.4+110831 & 34 Psc B          & 00:10:02.38 & +11:08:37.8 &  6.8 &
     8.65 $\pm$ 0.02 &  8.34 $\pm$ 0.03 &  8.18 $\pm$ 0.02 & & G1 & - \\
    \noalign{\smallskip}
    J001051.6+105140 & -                 & 00:10:51.39 & +10:51:40.2 &  3.4 &
    12.47 $\pm$ 0.02 & 11.96 $\pm$ 0.03 & 11.85 $\pm$ 0.02 & & K2 & K2 \\
    \noalign{\smallskip}
    J001749.7+161952 & 39 Psc            & 00:17:49.97 & +16:19:51.8 &  3.5 &
     6.29 $\pm$ 0.02 &  6.09 $\pm$ 0.02 &  6.05 $\pm$ 0.02 & & F4 & - \\
    \noalign{\smallskip}
    J002953.1+044524 & -                 & 00:29:53.22 & +04:45:23.3 &  1.4 &
    10.14 $\pm$ 0.04 &  9.88 $\pm$ 0.04 &  9.80 $\pm$ 0.04 & & F6 & - \\
    \noalign{\smallskip}
    J005822.9$-$274016 & Bok II 284        & 00:58:22.98 & -27:40:14.0 &  2.2 &
    10.84 $\pm$ 0.03 & 10.35 $\pm$ 0.03 & 10.24 $\pm$ 0.02 & & K1 & - \\
    \noalign{\smallskip}
    J012757.2+190000 & -                 & 01:27:57.15 & +19:00:00.9 &  1.4 &
     9.82 $\pm$ 0.02 &  9.31 $\pm$ 0.03 &  9.19 $\pm$ 0.02 & & K2 & - \\
    \noalign{\smallskip}
    J012757.3+185923 & BD+18 193         & 01:27:57.37 & +18:59:25.0 &  1.1 &
     8.24 $\pm$ 0.02 &  7.94 $\pm$ 0.03 &  7.86 $\pm$ 0.02 & & G0 & - \\
    \noalign{\smallskip}
    J014100.6$-$675328$^\dag$
                       & BL Hyi            & 01:41:00.38 & -67:53:27.3 &  2.2 &
    14.32 $\pm$ 0.04 & 13.66 $\pm$ 0.04 & 13.25 $\pm$ 0.05 & & M6 & Am\\
    \noalign{\smallskip}
    J021830.0$-$045514 & -                 & 02:18:29.91 & -04:55:13.7 &  1.6 &
    10.92 $\pm$ 0.02 & 10.28 $\pm$ 0.02 & 10.06 $\pm$ 0.02 & & M1 & - \\
    \noalign{\smallskip}
    J040744.6$-$710846 & -                 & 04:07:44.64 & -71:08:47.0 &  0.2 &
    14.11 $\pm$ 0.03 & 13.53 $\pm$ 0.04 & 13.25 $\pm$ 0.04 & & M4 & M2$-$3e\\
    \noalign{\smallskip}
    J040807.2$-$712702 & EXO 0408.4-7134   & 04:08:07.07 & -71:27:00.9 &  1.7 &
     9.66 $\pm$ 0.02 &  9.00 $\pm$ 0.02 &  8.84 $\pm$ 0.03 & & K8 & - \\
    \noalign{\smallskip}
    J051617.1+794408 & HD 32558          & 05:16:17.33 & +79:44:10.8 &  2.3 &
     9.55 $\pm$ 0.11 &  9.61 $\pm$ 0.08 &  9.21 $\pm$ 0.04 & & A7 & - \\
    \noalign{\smallskip}
    J052048.9$-$454128 & HD 274404         & 05:20:49.33 & -45:41:30.0 &  4.8 &
     9.69 $\pm$ 0.03 &  9.12 $\pm$ 0.02 &  8.96 $\pm$ 0.02 & & K4 & - \\
    \noalign{\smallskip}
    J052155.0$-$252200 & -                 & 05:21:55.21 & -25:22:00.7 &  2.6 &
    11.78 $\pm$ 0.03 & 11.42 $\pm$ 0.02 & 11.32 $\pm$ 0.02 & & G6 & - \\
    \noalign{\smallskip}
    J062425.7$-$642958 & GSC 08902-01379   & 06:24:25.81 & -64:29:57.0 &  1.4 &
    10.21 $\pm$ 0.02 &  9.89 $\pm$ 0.02 &  9.80 $\pm$ 0.02 & & G2 & - \\
    \noalign{\smallskip}
    J074359.7+744057 & -                 & 07:44:00.60 & +74:40:56.4 &  3.7 &
    11.89 $\pm$ 0.02 & 11.50 $\pm$ 0.02 & 11.41 $\pm$ 0.02 & & K0 & K5 \\
    \noalign{\smallskip}
    J080309.8+650807 & HD 65497          & 08:03:09.50 & +65:08:08.8 &  2.5 &
     6.90 $\pm$ 0.02 &  6.80 $\pm$ 0.02 &  6.74 $\pm$ 0.02 & & A8 & - \\
    \noalign{\smallskip}
    J085427.8+584158 & GSC 03811-00428   & 08:54:27.81 & +58:42:02.0 &  3.2 &
     8.38 $\pm$ 0.02 &  8.00 $\pm$ 0.04 &  7.88 $\pm$ 0.02 & & G6 & - \\
    \noalign{\smallskip}
    J091043.4+054757 & -                 & 09:10:43.31 & +05:48:01.0 &  3.6 &
    12.82 $\pm$ 0.03 & 12.29 $\pm$ 0.03 & 11.95 $\pm$ 0.03 & & M5 & M5 \\
    \noalign{\smallskip}
    J095955.2+251549 & -                 & 09:59:55.04 & +25:15:51.0 &  3.3 &
    10.85 $\pm$ 0.02 & 10.60 $\pm$ 0.02 & 10.59 $\pm$ 0.02 & & F6 & A9 \\
    \noalign{\smallskip}
    J102044.1+081424 & -                 & 10:20:44.06 & +08:14:23.4 &  1.5 &
    10.35 $\pm$ 0.02 &  9.76 $\pm$ 0.02 &  9.47 $\pm$ 0.02 & & M4 & M5e \\
    \noalign{\smallskip}
    J105131.1+573439 & [IUY2001] 2       & 10:51:30.93 & +57:34:39.5 &  1.5 &
    11.04 $\pm$ 0.02 & 10.47 $\pm$ 0.03 & 10.21 $\pm$ 0.02 & & M3 & M3$-$4e\\
    \noalign{\smallskip}
    J110320.1+355803 & HD 95735          & 11:03:20.23 & +35:58:11.7 &  8.2 &
     4.20 $\pm$ 0.24 &  3.64 $\pm$ 0.20 &  3.25 $\pm$ 0.31 & & M6: & M2 \\
    \noalign{\smallskip}
    J122655.1+012002 & -                 & 12:26:54.97 & +01:20:00.3 &  3.1 &
    14.39 $\pm$ 0.03 & 13.80 $\pm$ 0.03 & 13.57 $\pm$ 0.05 & & M2 & M1$-$3\\
    \noalign{\smallskip}
    J122751.2+333842 & -                & 12:27:51.19 & +33:38:43.3 &  0.9 &
     9.44 $\pm$ 0.02 &  8.82 $\pm$ 0.04 &  8.64 $\pm$ 0.02 & & M0 & K7 \\
    \noalign{\smallskip}
    J122837.3+015720 & GSC 00282-00187  & 12:28:37.19 & +01:57:20.4 &  2.0 &
    10.90 $\pm$ 0.03 & 10.26 $\pm$ 0.03 & 10.04 $\pm$ 0.02 & & M2 & - \\
    \noalign{\smallskip}
    J122942.3+015525 & GSC 00282-00477  & 12:29:42.47 & +01:55:24.8 &  1.6 &
    11.39 $\pm$ 0.02 & 10.98 $\pm$ 0.02 & 10.89 $\pm$ 0.03 & & G9 & F9$-$G2 \\
    \noalign{\smallskip}
    J123208.7+640304 & -                & 12:32:08.90 & +64:03:02.8 &  1.8 &
    12.73 $\pm$ 0.02 & 12.13 $\pm$ 0.02 & 11.95 $\pm$ 0.02 & & K7 & K7 \\
    \noalign{\smallskip}
    J123549.1$-$395026 & -                & 12:35:48.93 & -39:50:24.5 &  2.9 &
     9.79 $\pm$ 0.03 &  9.22 $\pm$ 0.02 &  8.94 $\pm$ 0.02 & & M4 & - \\
    \noalign{\smallskip}
    J123600.7$-$395217 & CD-39 7717B      & 12:36:00.55 & -39:52:15.6 &  3.1 &
     9.15 $\pm$ 0.02 &  8.53 $\pm$ 0.04 &  8.35 $\pm$ 0.03 & & K7 & - \\
    \noalign{\smallskip}
    J124938.7$-$060444 & IM Vir           & 12:49:38.68 & -06:04:44.7 &  0.7 &
     8.20 $\pm$ 0.02 &  7.74 $\pm$ 0.02 &  7.67 $\pm$ 0.02 & & K0 & - \\
    \noalign{\smallskip}
    J133321.2+503102 & GSC 03469-00261  & 13:33:21.00 & +50:31:03.7 &  2.9 &
    10.17 $\pm$ 0.02 &  9.90 $\pm$ 0.03 &  9.88 $\pm$ 0.03 & & F7 & F6: \\
    \noalign{\smallskip}
    J133626.9$-$342636 & -                & 13:36:26.90 & -34:26:36.5 &  0.5 &
    11.85 $\pm$ 0.04 & 11.36 $\pm$ 0.04 & 11.21 $\pm$ 0.03 & & K2 & - \\
    \noalign{\smallskip}
    J134732.0+582103 & -                & 13:47:31.95 & +58:21:03.5 &  0.5 &
    11.24 $\pm$ 0.02 & 10.60 $\pm$ 0.02 & 10.41 $\pm$ 0.02 & & K9 & M3e \\
    \noalign{\smallskip}
    J140219.6$-$110458 & HD 122473        & 14:02:19.63 & -11:04:58.3 &  0.6 &
     6.76 $\pm$ 0.02 &  6.32 $\pm$ 0.03 &  6.22 $\pm$ 0.02 & & K0 & - \\
    \noalign{\smallskip}
    J140936.9+261632 & -                & 14:09:36.86 & +26:16:31.9 &  0.7 &
    11.18 $\pm$ 0.02 & 10.55 $\pm$ 0.03 & 10.35 $\pm$ 0.02 & & M1 & M3e \\
    \noalign{\smallskip}
    J142800.1+424409 & -                & 14:28:00.01 & +42:44:10.7 &  1.6 &
    11.82 $\pm$ 0.02 & 11.20 $\pm$ 0.02 & 10.98 $\pm$ 0.02 & & M2 & - \\
    \noalign{\smallskip}
    J142901.2+423048 & HD 127244        & 14:29:01.32 & +42:30:50.6 &  2.0 &
     7.73 $\pm$ 0.02 &  7.41 $\pm$ 0.02 &  7.33 $\pm$ 0.02 & & G4 & - \\
    \noalign{\smallskip}
    J143923.1+640912 & HD 129390        & 14:39:22.83 & +64:09:14.0 &  2.7 &
     6.72 $\pm$ 0.02 &  6.52 $\pm$ 0.02 &  6.49 $\pm$ 0.02 & & F4 & - \\
    \noalign{\smallskip}
    J153156.6$-$082610 & HD 138372        & 15:31:56.45 & -08:26:09.7 &  2.9 &
     7.51 $\pm$ 0.02 &  7.01 $\pm$ 0.04 &  6.87 $\pm$ 0.03 & & K2 & - \\
    \noalign{\smallskip}
    J162911.1+780442$^\dag$
                       & WD 1631+78       & 16:29:10.31 & +78:04:39.9 &  3.9 &
    10.97 $\pm$ 0.02 & 10.40 $\pm$ 0.02 & 10.16 $\pm$ 0.01 & & M2 & WD \\
    \noalign{\smallskip}
    J162944.8+781128 & RIXOS 122-10     & 16:29:44.55 & +78:11:27.8 &  1.1 &
    12.36 $\pm$ 0.02 & 11.77 $\pm$ 0.02 & 11.54 $\pm$ 0.02 & & M2 & M4-5e \\
    \noalign{\smallskip}
    J165313.3+021645 & -                & 16:53:13.28 & +02:16:45.9 &  0.7 &
    10.19 $\pm$ 0.02 &  9.94 $\pm$ 0.03 &  9.88 $\pm$ 0.02 & & F6 & - \\
    \noalign{\smallskip}
    J165710.5+352024 & -                & 16:57:10.47 & +35:20:25.0 &  1.0 &
    11.80 $\pm$ 0.02 & 11.22 $\pm$ 0.01 & 11.11 $\pm$ 0.02 & & K4 & K4e \\
    \noalign{\smallskip}
    J205847.0$-$423704 & -                & 20:58:47.02 & -42:37:04.4 &  0.4 &
    10.71 $\pm$ 0.02 & 10.27 $\pm$ 0.03 & 10.22 $\pm$ 0.02 & & G8 & - \\
    \noalign{\smallskip}
    J212635.8$-$445046 & HD 203915        & 21:26:35.85 & -44:50:47.7 &  1.7 &
     6.98 $\pm$ 0.02 &  6.76 $\pm$ 0.03 &  6.70 $\pm$ 0.02 & & F5 & - \\
    \noalign{\smallskip}
    J213840.5$-$424241 & HD 205756        & 21:38:40.55 & -42:42:39.3 &  2.0 &
     8.17 $\pm$ 0.02 &  7.97 $\pm$ 0.06 &  7.85 $\pm$ 0.03 & & F4 & - \\
    \noalign{\smallskip}
    J215323.7+173018 & -                & 21:53:23.66 & +17:30:20.1 &  1.7 &
    11.13 $\pm$ 0.02 & 10.58 $\pm$ 0.02 & 10.48 $\pm$ 0.02 & & K3 & K2 \\
    \noalign{\smallskip}
    J221750.4$-$083210 & -                & 22:17:50.40 & -08:32:10.2 &  0.6 &
    11.98 $\pm$ 0.02 & 11.38 $\pm$ 0.02 & 11.22 $\pm$ 0.02 & & K8 & M0e \\
    \noalign{\smallskip}
    J222852.2$-$050915 & HD 213039        & 22:28:52.40 & -05:09:13.3 &  2.9 &
     8.59 $\pm$ 0.02 &  8.41 $\pm$ 0.04 &  8.31 $\pm$ 0.02 & & F3 & - \\
    \noalign{\smallskip}
    J224833.3$-$511900 & $\epsilon$ Gru   & 22:48:33.29 & -51:19:00.6 &  0.9 &
     3.18 $\pm$ 0.28 &  3.16 $\pm$ 0.27 &  3.19 $\pm$ 0.36 & & B9 & - \\
    \noalign{\smallskip}
    J224846.6$-$505929 & -                & 22:48:46.62 & -50:59:28.1 &  1.5 &
    11.72 $\pm$ 0.02 & 11.28 $\pm$ 0.02 & 11.15 $\pm$ 0.02 & & K0 & - \\
    \noalign{\smallskip}
    J225349.6$-$172137 & -                & 22:53:49.69 & -17:21:35.8 &  1.6 &
    11.19 $\pm$ 0.03 & 10.62 $\pm$ 0.02 & 10.32 $\pm$ 0.02 & & M4 & M1-2 \\
    \noalign{\smallskip}
    J230408.2+031820 & & 23:04:08.37 & +03:18:21.4 &  1.5 &
     9.57 $\pm$ 0.03 &  8.93 $\pm$ 0.04 &  8.85 $\pm$ 0.03 & & K7 & - \\
    \noalign{\smallskip}
    J231541.2$-$424125 & HD 219369        & 23:15:41.42 & -42:41:25.9 &  1.8 &
     8.88 $\pm$ 0.03 &  8.67 $\pm$ 0.04 &  8.56 $\pm$ 0.02 & & F4 & - \\
    \noalign{\smallskip}
    J231553.0$-$423800 & -                & 23:15:52.96 & -42:38:00.0 &  0.6 &
    10.03 $\pm$ 0.02 &  9.52 $\pm$ 0.03 &  9.40 $\pm$ 0.02 & & K2 & - \\
    \noalign{\smallskip}
    J233325.7$-$152240 & -                & 23:33:25.99 & -15:22:38.1 &  3.6 &
    10.82 $\pm$ 0.02 & 10.26 $\pm$ 0.02 & 10.12 $\pm$ 0.02 & & K3 & K4 \\
    \noalign{\smallskip}
    J235032.3+363156 & -                & 23:50:32.36 & +36:31:59.5 &  2.7 &
    11.53 $\pm$ 0.03 & 11.04 $\pm$ 0.03 & 10.91 $\pm$ 0.02 & & K1 & K1 \\
    \noalign{\smallskip}
    \end{longtable}
\end{landscape}

\small
    \begin{longtable}{lcccccccc}
    \caption{\label{tab:fit}
             Best fit parameters derived from the spectral analysis
             of the 58 stellar sources in the XMM-Newton Bright Serendipitous
             survey. Errors are measured as the 90\% confidence range.}\\
    \hline\hline
    \noalign{\smallskip}
    Id. & $kT_{1}$ & $kT_{2}$ & $EM_{1}$/$EM_{2}$ & 
    $Z/Z_\odot$ & $N_\mathrm{H}$ & $\chi^2$ & 
    counts$^\dag$ & Unabsorbed flux$^\ddag$ \\
    \noalign{\smallskip}
    (XBSS $\dots$) & (keV) & (keV) & & &
    ($\times 10^{21}$ cm$^{-2}$) & (d.o.f.) & &
    ($\times 10^{-13}$ erg cm$^{-2}$ s$^{-1}$)\\
    \noalign{\smallskip}
    \hline
    \endfirsthead
    \caption{continued.}\\
    \hline\hline
    \noalign{\smallskip}
    Id. & $kT_{1}$ & $kT_{2}$ & $EM_{1}$/$EM_{2}$ & 
    $Z/Z_\odot$ & $N_\mathrm{H}$ & $\chi^2$ & 
    counts$^\dag$ & Unabsorbed flux$^\ddag$ \\
    \noalign{\smallskip}
    (XBSS $\dots$) & (keV) & (keV) & & & 
    ($\times 10^{21}$ cm$^{-2}$) & (d.o.f.) &  &
    ($\times 10^{-13}$ erg cm$^{-2}$ s$^{-1}$)\\
    \noalign{\smallskip}
    \hline
    \endhead
    \noalign{\smallskip}
    \hline
    \noalign{\smallskip}
    \endfoot
    \noalign{\smallskip}
    J001002.4+110831 & 0.98$_{-0.11}^{+0.09}$ & -                      &
    -    & 0.09$_{-0.04}^{+0.05}$ & 0.34$_{-0.34}^{+0.55}$ & 1.25 (66) &
    828 & 7.31$_{-1.01}^{+1.13}$ \\
    \noalign{\smallskip}
    J001051.6+105140$^{**}$
                     & 0.10$_{-0.02}^{+0.01}$ & 2.01$_{-0.33}^{+0.62}$ &
    6.60 & 1.33$_{-0.36}^{+0.38}$ & 2.12$_{-0.40}^{+0.59}$ & 0.69 (24) &
    69 & 0.84$_{-0.00}^{+0.00}$ \\
    \noalign{\smallskip}
    J001749.7+161952 & 0.41$_{-0.03}^{+0.04}$ & -                      &
    -    & 1.09$_{-0.08}^{+0.08}$ & 0.00$_{-0.00}^{+0.33}$ & 0.85 (53) &
    890 & 1.42$_{-1.26}^{+0.14}$ \\
    \noalign{\smallskip}
    J002953.1+044524 & 0.23$_{-0.04}^{+0.04}$ & -                      &
    -    & 0.94$_{-0.12}^{+0.13}$ & 5.89$_{-1.01}^{+1.03}$ & 1.13 (29) &
    361 & 9.25$_{-8.40}^{+2.38}$ \\
    \noalign{\smallskip}
    J005822.9$-$274016 & 0.81$_{-0.23}^{+0.19}$ & -                      &
    -    & 0.04$_{-0.03}^{+0.04}$ & 0.13$_{-0.13}^{+1.11}$ & 0.81 (36) &
    434 & 1.35$_{-0.41}^{+0.32}$ \\
    \noalign{\smallskip}
    J012757.2+190000 & 1.00$_{-0.13}^{+0.10}$ & -                      &
    -    & 0.14$_{-0.07}^{+0.11}$ & 0.00$_{-0.00}^{+0.57}$ & 1.13 (31) &
    391 & 5.94$_{-2.52}^{+0.15}$ \\
    \noalign{\smallskip}
    J012757.3+185923 & 0.75$_{-0.05}^{+0.04}$ & -                      &
    -    & 0.16$_{-0.05}^{+0.07}$ & 0.25$_{-0.25}^{+0.50}$ & 0.94 (81) &
    1086 & 18.73$_{-3.18}^{+1.86}$ \\
    \noalign{\smallskip}
    J014100.6$-$675328 & 10.19$_{-1.08}^{+1.60}$ & -                     &
    -    & 1.14$_{-0.36}^{+0.39}$ & 0.00$_{-0.00}^{+0.03}$ & 1.18 (328)&
    9567 & 8.01$_{-1.48}^{+2.75}$ \\
    \noalign{\smallskip}
    J021830.0$-$045514 & 0.34$_{-0.06}^{+0.05}$ & 1.16$_{-0.02}^{+0.02}$ &
    1.27 & 0.16$_{-0.03}^{+0.04}$ & 0.30$_{-0.25}^{+0.29}$ & 1.21 (226) &
    6044 & 2.18$_{-0.18}^{+0.34}$ \\
    \noalign{\smallskip}
    J040744.6$-$710846 & 0.99$_{-0.16}^{+0.35}$ & -                     &
    -    & 0.06$_{-0.04}^{+0.07}$ & 0.05$_{-0.05}^{+0.05}$ & 1.16 (39) &
    323 & 0.70$_{-0.22}^{+0.19}$ \\
    \noalign{\smallskip}
    J040807.2$-$712702 & 0.35$_{-0.04}^{+0.07}$ & 0.95$_{-0.06}^{+0.07}$ &
    1.29 & 0.11$_{-0.02}^{+0.03}$ & 0.19$_{-0.17}^{+0.36}$ & 1.00 (215) &
    3222 & 3.97$_{-0.53}^{+0.61}$ \\
    \noalign{\smallskip}
    J051617.1+794408$^{*}$
                       & 0.34$_{-0.02}^{+0.03}$ & 0.92$_{-0.05}^{+0.05}$ &
    0.91 & 0.13$_{-0.02}^{+0.04}$ & 0.10$_{-0.10}^{+0.24}$ & 0.91 (332) & 
    5170 & 2.42$_{-0.73}^{+0.05}$ \\
    \noalign{\smallskip}
    J052048.9$-$454128 & 0.40$_{-0.06}^{+0.07}$ & 0.93$_{-0.09}^{+0.09}$ &
    0.77 & 0.11$_{-0.04}^{+0.06}$ & 0.00$_{-0.00}^{+0.34}$ & 0.94 (263) &
    3708 & 4.47$_{-0.33}^{+0.85}$ \\
    \noalign{\smallskip}
    J052155.0$-$252200 & 0.72$_{-0.13}^{+0.10}$ & -                      &
    -    & 0.23$_{-0.14}^{+0.70}$ & 3.37$_{-1.16}^{+1.22}$ & 0.75 (54) &
    220 & 1.31$_{-0.99}^{+0.27}$ \\
    \noalign{\smallskip}
    J062425.7$-$642958 & 0.47$_{-0.19}^{+0.11}$ & -                      &
    -    & 0.11$_{-0.08}^{+0.54}$ & 1.34$_{-1.34}^{+3.21}$ & 0.66 (35) &
    227 & 1.73$_{-1.14}^{+0.37}$ \\
    \noalign{\smallskip}
    J074359.7+744057 & 0.30$_{-0.04}^{+0.05}$ & 1.20$_{-0.20}^{+0.40}$ &
    5.43 & 0.08$_{-0.06}^{+0.10}$ & 2.73$_{-1.09}^{+1.06}$ & 0.92 (65) &
    1430 & 4.03$_{-3.53}^{+0.01}$ \\
    \noalign{\smallskip}
    J080309.8+650807 & 0.50$_{-0.03}^{+0.05}$ & -                      &
    -    & 1.45$_{-0.14}^{+0.71}$ & 0.00$_{-0.00}^{+0.80}$ & 1.40 (59) &
    707 & 0.96$_{-0.48}^{+0.13}$ \\
    \noalign{\smallskip}
    J085427.8+584158 & 0.61$_{-0.08}^{+0.06}$ & -                      &
    -    & 0.09$_{-0.04}^{+0.06}$ & 0.00$_{-0.00}^{+0.52}$ & 0.71 (62) &
    945 & 1.36$_{-0.35}^{+0.21}$ \\
    \noalign{\smallskip}
    J091043.4+054757 & 1.02$_{-0.22}^{+0.26}$ & -                      &
    -    & 0.04$_{-0.03}^{+0.05}$ & 0.21$_{-0.21}^{+0.73}$ & 1.05 (74) &
    1235 & 1.18$_{-0.23}^{+0.26}$ \\
    \noalign{\smallskip}
    J095955.2+251549 & 1.01$_{-0.06}^{+0.05}$ & -                      &
    -    & 0.12$_{-0.03}^{+0.04}$ & 0.76$_{-0.35}^{+0.21}$ & 1.05 (114) &
    3214 & 1.39$_{-0.23}^{+0.21}$ \\
    \noalign{\smallskip}
    J102044.1+081424 & 0.33$_{-0.03}^{+0.04}$ & -                      &
    -    & 0.03$_{-0.02}^{+0.03}$ & 1.33$_{-0.46}^{+0.45}$ & 1.15 (48) &
    1117 & 2.72$_{-1.58}^{+0.34}$ \\
    \noalign{\smallskip}
    J105131.1+573439 & 0.32$_{-0.08}^{+0.08}$ & 0.93$_{-0.14}^{+0.18}$ &
    1.14 & 0.15$_{-0.06}^{+0.20}$ & 0.05$_{-0.05}^{+0.90}$ & 0.94 (79) &
    1834 & 1.02$_{-0.53}^{+0.10}$ \\
    \noalign{\smallskip}
    J110320.1+355803 & 0.27$_{-0.02}^{+0.02}$ & 0.71$_{-0.06}^{+0.13}$ &
    4.18 & 0.06$_{-0.01}^{+0.01}$ & 0.00$_{-0.00}^{+0.01}$ & 1.29 (391) &
    11147 & 3.97$_{-2.03}^{+1.92}$\\
    \noalign{\smallskip}
    J122751.2+333842 & 0.26$_{-0.03}^{+0.04}$ & 0.89$_{-0.08}^{+0.09}$ &
    1.23 & 0.21$_{-0.09}^{+0.26}$ & 0.00$_{-0.00}^{+0.71}$ & 0.70 (67) &
    1844 & 1.01$_{-0.77}^{+0.03}$ \\
    \noalign{\smallskip}
    J122655.1+012002 & 1.08$_{-0.31}^{+0.31}$ & -                      &
    -    & 0.08$_{-0.07}^{+0.16}$ & 0.00$_{-0.00}^{+0.11}$ & 0.80 (59) &
    708 & 0.33$_{-0.26}^{+0.09}$ \\
    \noalign{\smallskip}
    J122837.3+015720 & 0.27$_{-0.04}^{+0.03}$ & 1.04$_{-0.08}^{+0.10}$ &
    1.99 & 0.17$_{-0.06}^{+0.08}$ & 1.12$_{-0.74}^{+1.30}$ & 0.81 (135) &
    1857 & 1.78$_{-0.83}^{+0.34}$ \\
    \noalign{\smallskip}
    J122942.3+015525 & 0.29$_{-0.06}^{+0.07}$ & 1.03$_{-0.10}^{+0.13}$ &
    0.90 & 0.35$_{-0.14}^{+0.27}$ & 1.94$_{-1.13}^{+1.20}$ & 0.96 (79) &
    1059 & 1.72$_{-0.78}^{+0.33}$ \\
    \noalign{\smallskip}
    J123208.7+640304 & 0.25$_{-0.07}^{+0.10}$ & 1.09$_{-0.15}^{+0.23}$ &
    1.09 & 0.10$_{-0.07}^{+0.24}$ & 0.28$_{-0.28}^{+1.95}$ & 0.85 (44) &
    534 & 0.64$_{-0.40}^{+0.14}$\\
    \noalign{\smallskip}
    J123549.1$-$395026 & 0.32$_{-0.01}^{+0.01}$ & 0.90$_{-0.04}^{+0.04}$ &
    1.73 & 0.19$_{-0.04}^{+0.04}$ & 0.00$_{-0.00}^{+0.27}$ & 1.08 (216) &
    6004 & 2.16$_{-0.58}^{+0.19}$ \\
    \noalign{\smallskip}
    J123600.7$-$395217$^{*}$ 
                       & 0.31$_{-0.01}^{+0.01}$ & 1.01$_{-0.03}^{+0.02}$ &
    0.81 & 0.22$_{-0.04}^{+0.04}$ & 0.00$_{-0.00}^{+0.01}$ & 1.42 (617) & 
    34052 & 14.05$_{-3.64}^{+2.45}$ \\
    \noalign{\smallskip}
    J124938.7$-$060444 & 0.64$_{-0.01}^{+0.01}$ & 1.21$_{-0.02}^{+0.02}$ &
    0.83 & 0.17$_{-0.01}^{+0.01}$ & 0.01$_{-0.01}^{+0.02}$ & 1.15 (517) &
    40463 & 22.39$_{-1.46}^{+0.66}$ \\
    \noalign{\smallskip}
    J133321.2+503102 & 0.60$_{-0.04}^{+0.04}$ & -                      &
    -    & 0.09$_{-0.02}^{+0.03}$ & 0.98$_{-0.38}^{+0.41}$ & 1.09 (170) &
    1190 & 1.25$_{-0.20}^{+0.24}$ \\
    \noalign{\smallskip}
    J133626.9$-$342636 & 0.98$_{-0.41}^{+0.32}$ & -                      &
    -    & 0.07$_{-0.06}^{+0.15}$ & 0.03$_{-0.03}^{+2.15}$ & 0.48 (16) &
    466 & 0.51$_{-0.35}^{+0.15}$ \\
    \noalign{\smallskip}
    J134732.0+582103 & 0.82$_{-0.03}^{+0.03}$ & -                      &
    -    & 0.06$_{-0.01}^{+0.01}$ & 0.00$_{-0.00}^{+0.15}$ & 1.06 (177) &
    4120 & 1.75$_{-0.16}^{+0.15}$ \\
    \noalign{\smallskip}
    J140219.6$-$110458 & 0.44$_{-0.12}^{+0.16}$ & -                      &
    -    & 1.11$_{-0.94}^{+3.88}$ & 2.76$_{-1.68}^{+1.83}$ & 1.02 (39) &
    491 & 1.16$_{-0.62}^{+0.18}$ \\
    \noalign{\smallskip}
    J140936.9+261632 & 0.35$_{-0.06}^{+0.09}$ & 0.94$_{-0.18}^{+0.17}$ &
    2.48 & 0.08$_{-0.08}^{+0.06}$ & 0.29$_{-0.29}^{+1.03}$ & 1.04 (95) &
    2402 & 0.78$_{-0.29}^{+0.13}$ \\
    \noalign{\smallskip}
    J142800.1+424409 & 0.28$_{-0.05}^{+0.08}$ & 0.93$_{-0.11}^{+0.12}$ &
    1.87 & 0.23$_{-0.11}^{+0.36}$ & 1.00$_{-1.00}^{+2.07}$ & 0.84 (90) &
    1226 & 1.07$_{-0.40}^{+0.18}$ \\
    \noalign{\smallskip}
    J142901.2+423048 & 0.49$_{-0.08}^{+0.06}$ & -                      &
    -    & 0.99$_{-0.72}^{+4.00}$ & 0.00$_{-0.00}^{+0.42}$ & 1.15 (59) &
    805 & 0.70$_{-0.66}^{+0.50}$ \\
    \noalign{\smallskip}
    J143923.1+640912 & 0.49$_{-0.03}^{+0.02}$ & -                      &
    -    & 1.11$_{-0.62}^{+3.89}$ & 0.01$_{-0.01}^{+0.09}$ & 1.35 (120) &
    2155 & 1.61$_{-1.35}^{+0.26}$ \\
    \noalign{\smallskip}
    J153156.6$-$082610 & 0.57$_{-0.32}^{+0.08}$ & -                      &
    -    & 0.11$_{-0.04}^{+0.22}$ & 1.07$_{-0.98}^{+3.43}$ & 0.87 (84) &
    909 & 0.88$_{-0.82}^{+1.63}$ \\
    \noalign{\smallskip}
    J162911.1+780442 & 0.70$_{-0.41}^{+0.20}$ & -                      &
    -    & 0.04$_{-0.02}^{+0.02}$ & 0.00$_{-0.00}^{+0.92}$ & 1.00 (21) &
    280 & 1.13$_{-0.88}^{+0.08}$ \\
    \noalign{\smallskip}
    J162944.8+781128 & 0.74$_{-0.23}^{+0.16}$ & -                      &
    -    & 0.10$_{-0.08}^{+0.29}$ & 0.10$_{-0.10}^{+2.27}$ & 1.09 (9) & 
    150 & 0.80$_{-0.78}^{+0.19}$ \\
    \noalign{\smallskip}
    J165313.3+021645 & 0.64$_{-0.05}^{+0.04}$ & -                      &
    -    & 0.56$_{-0.30}^{+4.43}$ & 0.09$_{-0.09}^{+1.18}$ & 0.93 (79) &
    1318 & 0.70$_{-0.67}^{+0.15}$ \\
    \noalign{\smallskip}
    J165710.5+352024$^{**}$
                     & 0.85$_{-0.36}^{+0.28}$ & -                      &
    -    & 0.08$_{-0.07}^{+0.33}$ & 0.06$_{-0.06}^{+3.11}$ & 1.21 (7) &
    232 & 1.24$_{-0.60}^{+0.35}$ \\
    \noalign{\smallskip}
    J205847.0$-$423704 & 0.74$_{-0.09}^{+0.06}$ & -                      &
    -    & 0.14$_{-0.06}^{+0.07}$ & 0.34$_{-0.34}^{+0.57}$ & 1.21 (72) &
    1096 & 0.71$_{-0.31}^{+0.18}$ \\
    \noalign{\smallskip}
    J212635.8$-$445046 & 0.62$_{-0.06}^{+0.05}$ & -                      &
    -    & 0.11$_{-0.04}^{+0.07}$ & 0.17$_{-0.17}^{+0.78}$ & 1.14 (78) &
    1899 & 1.52$_{-0.70}^{+0.31}$ \\
    \noalign{\smallskip}
    J213840.5$-$424241 & 0.47$_{-0.14}^{+0.09}$ & -                      &
    -    & 0.97$_{-0.13}^{+0.13}$ & 0.00$_{-0.00}^{+1.93}$ & 0.97 (118) &
    1261 & 0.72$_{-0.67}^{+0.06}$ \\
    \noalign{\smallskip}
    J215323.7+173018 & 1.31$_{-0.29}^{+0.48}$ & -                      &
    -    & 0.14$_{-0.11}^{+0.30}$ & 1.05$_{-1.05}^{+1.36}$ & 0.98 (75) &
    507 & 0.69$_{-0.21}^{+0.27}$ \\
    \noalign{\smallskip}
    J221750.4$-$083210 & 0.23$_{-0.04}^{+0.05}$ & 1.09$_{-0.22}^{+0.28}$ &
    3.61 & 0.71$_{-0.12}^{+0.11}$ & 2.44$_{-1.59}^{+1.15}$ & 0.99 (49) &
    546 & 1.76$_{-1.43}^{+0.09}$ \\
    \noalign{\smallskip}
    J222852.2$-$050915 & 0.52$_{-0.07}^{+0.05}$ & -                      &
    -    & 0.50$_{-0.23}^{+1.92}$ & 0.36$_{-0.36}^{+0.88}$ & 1.04 (131) &
    1550 & 0.92$_{-0.62}^{+0.18}$ \\
    \noalign{\smallskip}
    J224833.3$-$511900 & 0.30$_{-0.03}^{+0.03}$ & -                      &
    -    & 0.65$_{-0.14}^{+0.13}$ & 4.94$_{-0.52}^{+0.73}$ & 1.03 (26) &
    170 & 2.41$_{-2.26}^{+0.64}$ \\
    \noalign{\smallskip}
    J224846.6$-$505929 & 0.69$_{-0.48}^{+0.14}$ & -                      &
    -    & 0.09$_{-0.07}^{+0.53}$ & 0.00$_{-0.00}^{+2.48}$ & 0.55 (20) &
    139 & 0.52$_{-0.39}^{+0.16}$ \\
    \noalign{\smallskip}
    J225349.6$-$172137$^{**}$ 
                     & 0.13$_{-0.06}^{+0.05}$ & -                      &
    -    & 4.74$_{-4.72}^{+0.26}$ & 8.88$_{-3.85}^{+8.46}$ & 0.94 (20) &
    71 & 61.89$_{-61.85}^{+2.00}$ \\
    \noalign{\smallskip}
    J230408.2+031820 & 0.31$_{-0.05}^{+0.09}$ & 0.91$_{-0.09}^{+0.13}$ &
    1.48 & 0.15$_{-0.05}^{+0.16}$ & 0.54$_{-0.12}^{+0.17}$ & 1.23 (81) &
    1936 & 4.17$_{-0.90}^{+0.69}$ \\
    \noalign{\smallskip}
    J231541.2$-$424125 & 0.47$_{-0.07}^{+0.07}$ & -                      &
    -    & 0.71$_{-0.53}^{+4.29}$ & 0.19$_{-0.19}^{+0.89}$ & 1.08 (79) &
    862 & 0.69$_{-0.63}^{+0.17}$ \\
    \noalign{\smallskip}
    J231553.0$-$423800 & 0.78$_{-0.17}^{+0.26}$ & -                      &
    -    & 0.02$_{-0.01}^{+0.02}$ & 0.92$_{-0.80}^{+0.79}$ & 1.01 (70) &
    1247 & 0.73$_{-0.21}^{+0.16}$ \\
    \noalign{\smallskip}
    J233325.7$-$152240 & 0.29$_{-0.07}^{+0.08}$ & 0.97$_{-0.02}^{+0.03}$ &
    2.41 & 0.09$_{-0.02}^{+0.05}$ & 1.09$_{-0.60}^{+0.54}$ & 1.31 (153) &
    2853 & 2.35$_{-0.91}^{+0.33}$ \\
    \noalign{\smallskip}
    J235032.3+363156 & 1.13$_{-0.14}^{+0.14}$ & -                      &
    -    & 0.34$_{-0.10}^{+0.09}$ & 0.00$_{-0.00}^{+0.52}$ & 0.77 (29) & 
    152 & 1.05$_{-0.64}^{+0.24}$ \\
    \noalign{\smallskip}
    \end{longtable}
    \noindent
    {\small $^{\dag}$ Total net counts in the EPIC (MOS + PN)
                      chips.\\ 
            $^{\ddag}$ Unabsorbed flux in 0.5--10 keV.\\
            $^{*}$ For these sources, a 3T-model has been 
                   used to obtain an accurate fitting to the hard tail of
                   their X-ray spectrum. The parameters of their third thermal
                   component are 
                   a) XBSS J051617.1+794408 (HD 32558):
                   $kT_{3} = 4.32_{-2.50}^{+3.28}$ keV, 
                   $EM_{1}/EM_{3} = 3.57$ and
                   b) XBSS J123600.7$-$395217 (CD-39 7717B, TWA 11B):
                   $kT_{3} = 2.39_{-0.15}^{+0.16}$ keV,
                   $EM_{1}/EM_{3} = 0.75$.\\
            $^{**}$ Problematic sources. The fitting in these stars
                    is much less accurate due to the few number of
                    counts in their spectra after background
                    subtraction.}


%
\begin{table*}[!h]
\caption{Unabsorbed X-ray luminosity for the stars of the sample 
         with known paralaxes. Emission measure ($EM_{1}$) is 
         also given. For those stars where more than one thermal
         component has been fitted to the X-ray spectrum, 
         $EM_{1}$ is the emission measure of the first component.}
\label{tab:selected}
\small
\begin{center}
\begin{tabular}{l r c c}
\hline\hline
Star & \multicolumn{1}{c}{distance} & $\log{EM_{1}}$ & $\log{L_\mathrm{X}}$ \\
(XBSS $\dots$) & \multicolumn{1}{c}{(pc)} & (cm$^{-3}$) & (erg s$^{-1}$) \\
\hline
J001002.4+110831   &  99.60 & 54.30 & 29.94 \\
J001749.7+161952   &  46.17 & 52.33 & 28.56 \\
J080309.8+650807   & 114.02 & 52.77 & 29.17 \\
J110320.1+355803   &   2.55 & 51.21 & 26.49 \\
J123600.7$-$395217 &  67.07 & 53.56 & 29.88 \\
J212635.8$-$445046 & 114.16 & 53.81 & 29.37 \\
J213840.5$-$424241 &  77.28 & 52.49 & 28.71 \\
J224833.3$-$511900 &  39.75 & 52.73 & 28.65 \\
J231541.2$-$424125 &  99.11 & 52.67 & 28.91 \\
\hline
\end{tabular}
\end{center}

\end{table*}

\end{document}